\documentclass[useAMS,usenatbib,usegraphicx]{mn2e}
\usepackage{graphicx}
\usepackage{color}

\title[Messier 5: II. Blazhko stars]{Long-term photometric monitoring of Messier 5 variables:\\ II. Blazhko stars}

\author[Jurcsik et al.]{J. Jurcsik$^{1}$\thanks{E-mail: jurcsik@konkoly.hu},
 B. Szeidl$^{1}$, C. Clement$^{2}$, Zs. Hurta$^{1}$ and M. Lovas$^{1}$ \\
$^{1}$Konkoly Observatory of the Hungarian Academy of Sciences, H-1525 Budapest PO Box 67, Hungary\\
$^{2}$Dept. of Astronomy, University Toronto, Ontario, M5S3H8, Canada}
\begin{document}

\date{Accepted 2010 Oct 4; 29  Received 2010 Sept 22; in original form 2010 Aug 3}

\pagerange{\pageref{firstpage}--\pageref{lastpage}} \pubyear{2010}

\maketitle

\label{firstpage}

\begin{abstract}
The light curves of 50 RRab (RR0) stars in M5 collected in Paper I are investigated to detect Blazhko modulation. 18 Blazhko stars are identified, and modulation is suspected in two additional cases. The mean pulsation period of Blazhko stars is 0.04 d shorter than the mean period of the entire RRab sample in M5. Among the RRab stars with period shorter than 0.55 d the incidence rate of the modulation is as high as 60 per cent. The mean $B-V$ colours of Blazhko stars overlap with the colours of first overtone RRc (RR1) pulsators. The mean $V$ magnitudes of Blazhko stars are on the average 0.05-mag fainter than those of the RRab stars with stable light curves. Blazhko stars tend to be situated close to the zero-age horizontal branch at the blue edge of the fundamental-mode instability strip in M5. We speculate that this specific location hints that the Blazhko effect may have an evolutionary connection with the mode switch from the fundamental to the overtone-mode pulsation.

\end{abstract}


\section {Introduction}
The light-curve modulation of RRab stars (Blazhko effect) is one of the most outstanding problems of stellar pulsation theory, which has remained unsolved despite continuous observational and theoretical efforts.
Recent, accurate, ground based and space observations \citep{mw2,aql,szabo} reveal more and more details of the snapshots of the modulations' properties. However, contrary to the remarkable stability of the light curves of non-modulated RRab stars, the modulation properties of Blazhko variables may show significant changes on time-scales of years and decades e.g. in the cases of RR Gem \citep{rrgII}, RR Lyr \citep{rrl} and XZ Cyg \citep{xzc}. The only way to study these long-term changes is to combine archival data with modern observations.
Globular clusters, which have been already observed regularly on a time base longer than a century, are ideal targets for following the variations of the modulation properties of Blazhko stars, provided that the modulation is strong enough to be detected in photographic data. 

An accurate knowledge of the Blazhko-star population in globular clusters also yields an important information for determining the incidence of this phenomenon. Recent surveys show that about half the RRab stars exhibit light-curve modulation in some samples \citep{stat,corot}, while previous estimates gave only 10--30 per cent.

M5, one of the closest globular clusters with about 100 RR Lyrae stars having been continuously observed during the 20$^{th}$ century, is one of the few possible targets for these studies. 
Blazhko stars are quite numerous in M5 based on previous photographic and CCD investigations.
Already \citet[hereafter Oo41 ]{oo} noted that the light curves of V2, V4, V14, V18, V27, V52, V63 and V72 in M5 were not stable; they showed irregular or RW Dra-type variability.\footnote{The star RW Dra, previously known as 87.1906 Dra \citep{sz01} was the first RR Lyrae star to have long-period phase oscillations identified \citep{bl07}. The phenomenon is now referred to as the Blazhko effect.}
Blazhko effect was also identified by \citet[hereafter KK71]{kk} in V2, V14, V18, V58 and V63.
Using photographic observations, \cite{g76,g80a,g80b} determined Blazhko modulation periods for V2, V14 and V63.
Based on their CCD observations, \citet[hereafter R96]{r96} and \citet[hereafter K00]{k00} noted light-curve variability of V2, V4, V5, V8, V14, V27, V56, V65, V89 and V97.

A comprehensive study of the Blazhko stars in M5 is, however, still missing. In \citet[Paper I]{oc} we collected all the photometric data of M5 variables in order to study the period changes of RR Lyrae stars. These data are  utilized here to detect and study Blazhko variables.

\begin{figure}
\includegraphics[angle=0,width=8.3cm]{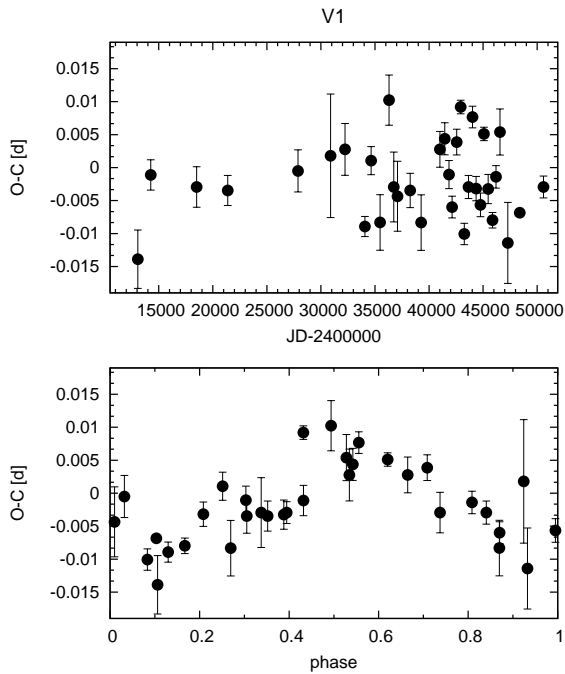}
\caption{Seasonal $O-C$ values of V1 (top panel) derived for the entire photometric data. Period analysis of the $O-C$ data  results in a 512 d periodicity of the phase variation. The bottom panel shows the $O-C$ data folded with this modulation period.} 
\label{v01oc}
\end{figure}

\begin{figure}
\includegraphics[angle=0,width=8.2cm]{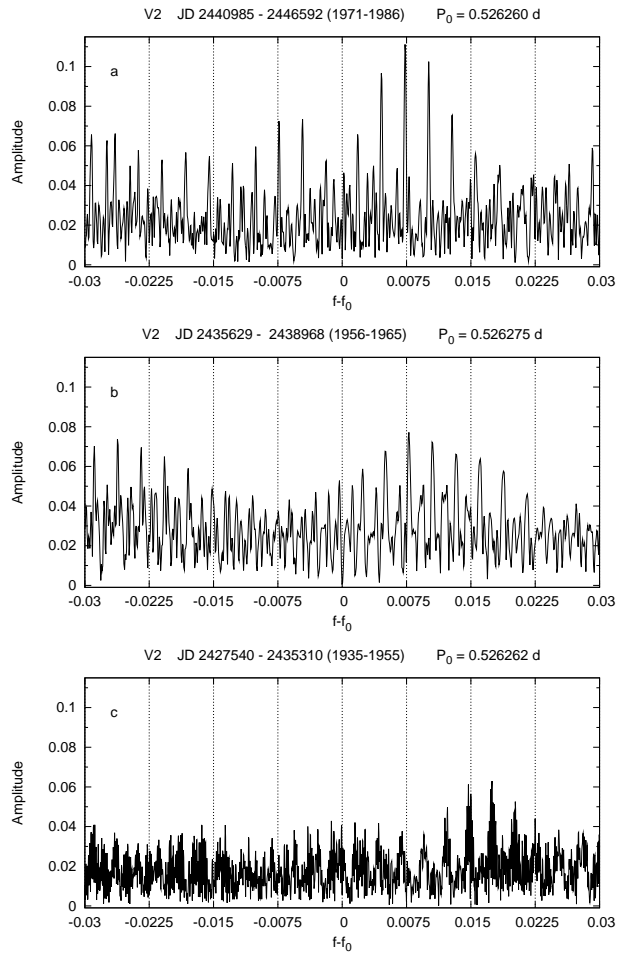}
\caption{The residual spectra in the vicinity of the main pulsation frequency, $f_0$, of different segments of V2's light curve are shown. The top ($a$), the middle ($b$) and the bottom ($c$) panels refer to data for the 1971--1986, 1956--1965 and 1934--1955 intervals, respectively. In panel $a$ frequencies at $\pm 0.0075$ cd$^{-1}$  appear, corresponding to a modulation with a 136-d period. In 1956--1965 this modulation could also be detected; however, the period was a bit shorter at that time, 130 d.  Some modulation features also appear in the earliest data (panel $c$), a signal at $f_0+0.018$ is the highest signal in the entire residual spectrum in the $0-10$ cd$^{-1}$ frequency range. The one-cycle-per-year alias peak of this frequency at $f_0+0.015$ is at a separation that is twice the separation of the modulation frequencies appearing in panels $a$~and~$b$. } 
\label{v02sp}
\end{figure}
\begin{figure}
\includegraphics[angle=0,width=8.2cm]{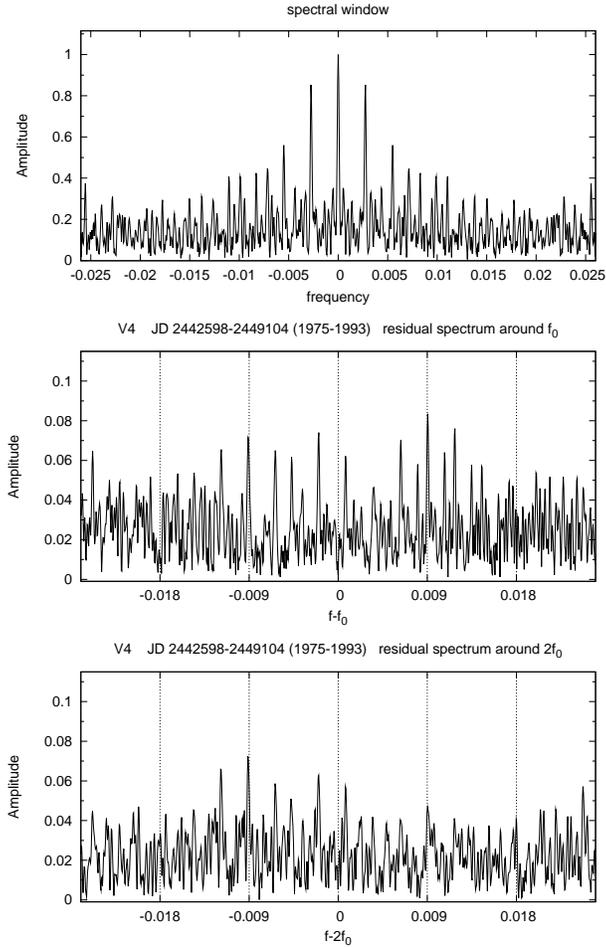}
\caption{The residual spectra of V4's photographic light curve between JD 2442553 and JD 2449104 are shown. The top panel shows the spectral window: yearly and monthly alias frequencies have significant amplitudes due to these periodicities appearing in the data distribution. The middle and the bottom panels show the amplitude spectra of the prewhitened data in the vicinity of $f_0$ and $2f_0$, respectively. Modulation frequencies appear at $f_0-f_{\mathrm{m}}$, $f_0+f_{\mathrm{m}}$ and $2f_0-f_{\mathrm{m}}$, $f_{\mathrm{m}}=0.009$ cd$^{-1}$,  corresponding to an $\approx110$ d modulation period.  } 
\label{v04sp}
\end{figure}
\begin{figure}
\includegraphics[angle=0,width=6.8cm]{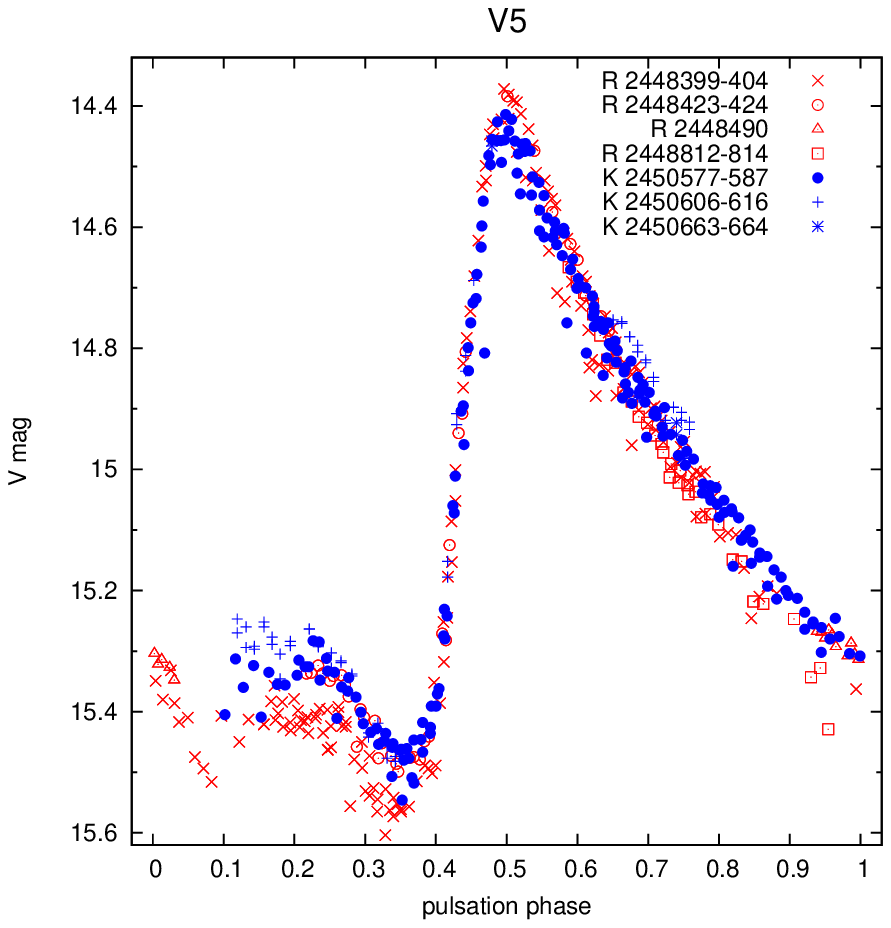}
\caption{CCD $V$ light curve of V5. The different parts of the data are plotted with different symbols as indicated by the key labels. The prefixes R and K refer to the R96 and K00 data, respectively.  }
\label{v5}
\end{figure}
\begin{figure}
\includegraphics[angle=0,width=8.2cm]{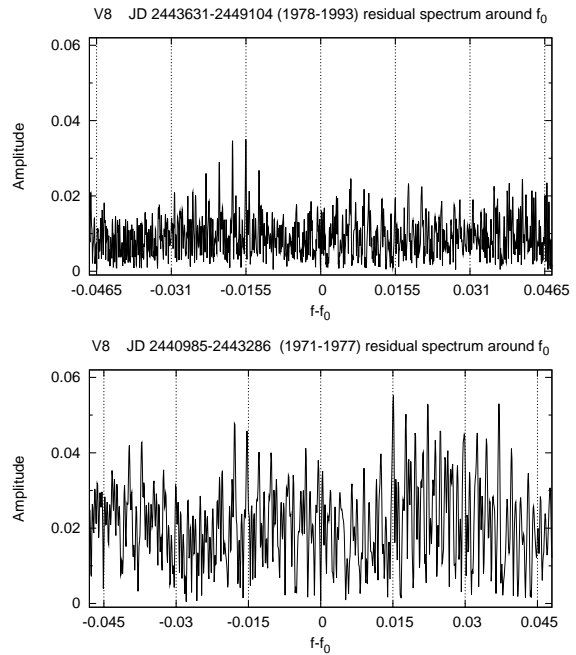}
\caption{The residual spectra of V8's photographic light curve for the 1978--1993 (top) and 1971--1977 (bottom) periods in the vicinity of $f_0$ are shown.  Modulation peaks with $f_{\mathrm{m}}=0.0155,0.0150$ cd$^{-1}$  ($\sim65$ d) separation appear in both data sets.  A yearly alias solution with 0.0183  cd$^{-1}$ (55 d) is also possible.} 
\label{v08sp}
\end{figure}
\begin{figure}
\includegraphics[angle=0,width=8.3cm]{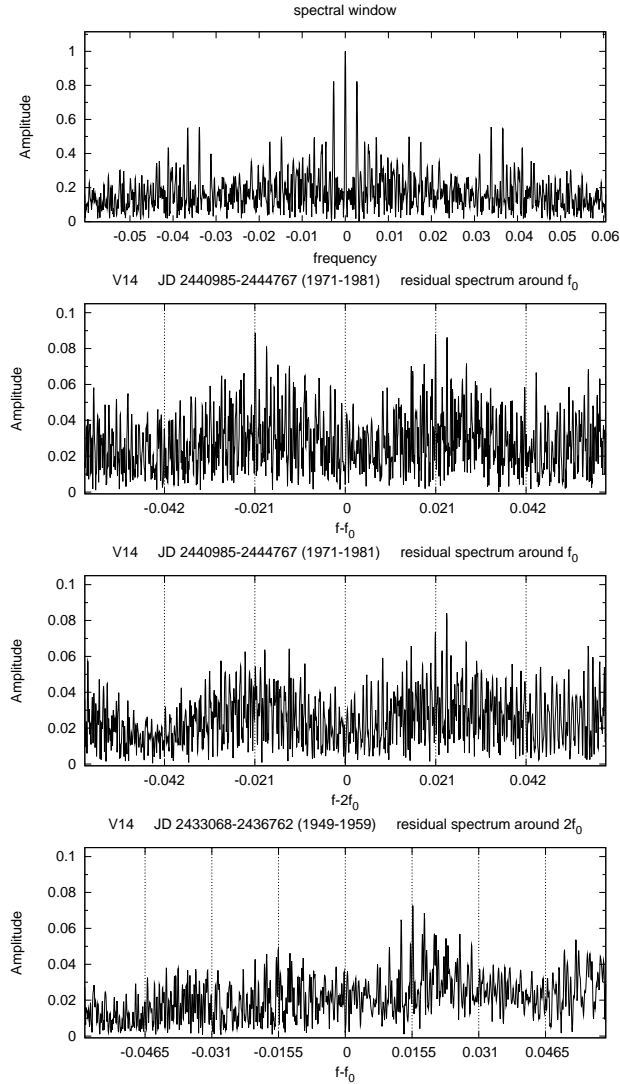}
\caption{The residual spectra of V14's photographic light curve between JD 2440985 and JD 2444767 are shown. The top panel shows the spectral window. The next two panels show the amplitude spectra of the prewhitened data in the vicinity of $f_0$ and $2f_0$, for this data set. Modulation frequencies appear in these panels at $f_0-f_{\mathrm{m}}$, $f_0+f_{\mathrm{m}}$ and $2f_0+f_{\mathrm{m}}$, $f_{\mathrm{m}}=0.021$ cd$^{-1}$, which corresponds to a 47.8-d modulation period. Bottom panel shows the residual amplitude spectrum at $f_0$ for another subset of the data, between JD 4233068 and JD 2436762. The modulation peak appearing in this data set corresponds to a significantly different modulation period, $P_{m}=63.8$ d.} 
\label{v14sp}
\end{figure}
\begin{figure}
\includegraphics[angle=-90,width=8.3cm]{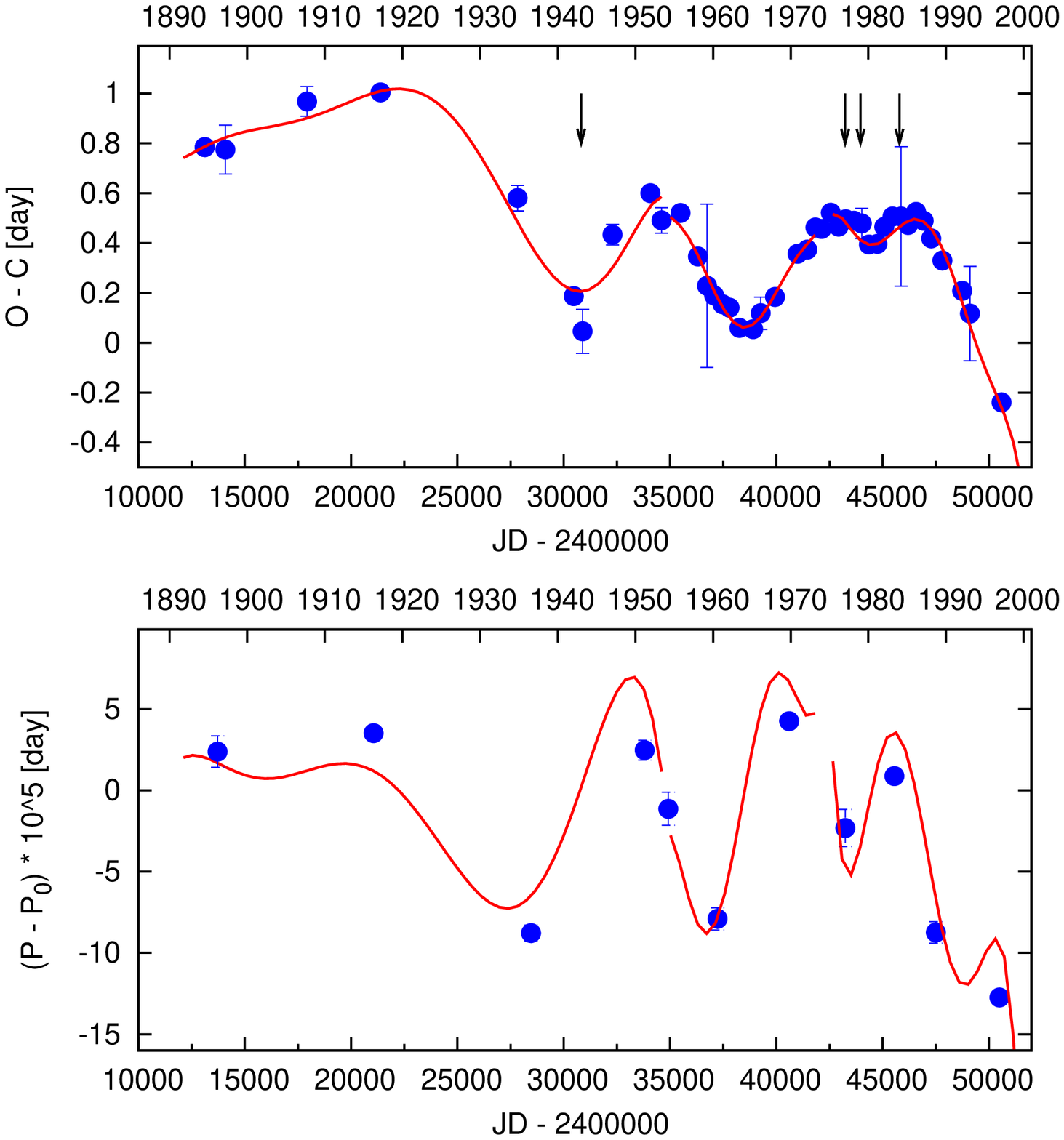}
\caption{Seasonal $O-C$ variation of V18 (top panel) and pulsation-period values derived for different subsets of the data. The drawn fits are the derivatives of the fits to the  $O-C$ data as described in Paper I. The arrows indicate the years when the pulsation amplitude was smaller than 0.5-mag. No connection between the small-amplitude phases and the period-change variation is evident.}
\label{v18oc}
\end{figure}

\section{Data and method}

The homogenized (see section 2.3 of Paper I) light curves of the RRab stars in M5 were carefully examined to identify variables showing light-curve instability.
Due to crowding problems, this was not possible for all the RRab stars in our sample. Variables lying in too crowded areas and/or having close bright companions could not be measured accurately enough to detect light-curve variability without doubt. This was especially true if no CCD observation of the variable was available. From the 65 RRab stars for which phase variation diagrams were constructed in Paper I, this was the case for 15 stars (V6, V13, V17, V25, V26, V36, V37, V54, V74, V83, V85, V90, V91, V92 and V96). These stars were denoted by the letter `$d$' in the `Remarks' column of Table 4 in Paper I.
The combined photographic and CCD $B$ light curves of the remaining 50 RRab stars were studied using different methods in order to detect light-curve variations and to determine modulation periods if it was possible. When CCD $V$ observations from different sources  were also available, these data were investigated separately. 

Different segments of the data were checked and analysed. If light-curve variability was suspected in any part of the data, Fourier analyses of the light curves and the maximum brightness/maximum timing data were also applied. The maximum brightness/phase data of globular-cluster variables are, however, very sparse (one-two data per season if any), and give useable information only in very few cases (e.g. in V19). 

We searched for modulation-frequency signals in the vicinity of 
the low-order pulsation frequency components ($kf_0, k<4$) 
corresponding to the same modulation frequency ($f_{\mathrm{m}}$). 
The modulation period ($P_{\mathrm{m}}=1/f_{\mathrm{m}}$) was then determined by a 
non-linear process that fitted the pulsation and some of the 
modulation frequencies ($kf_0, k<8$ and $kf_0\pm f_{\mathrm{m}}, k<4$) 
and their amplitudes and phases simultaneously to the data.

A Blazhko classification was assigned if at least one of the following criteria was fulfilled:

\begin{itemize}
\item{changes in the light curve's amplitude, shape and/or phase were detected,}
\item{at least three modulation frequencies, corresponding to the same modulation period, appeared around the frequencies of the pulsation components in the residual spectrum of one subset of the light-curve data, }
\item{frequencies, corresponding to the same modulation period, appeared in the residual spectra of different subsets of the observations.}

\end{itemize}

We note, however, that without accurate, extended, CCD photometric observations the results obtained from inhomogeneous, photographic data have to be regarded as somewhat uncertain. Even though the available CCD observations of the M5 variables show clear evidence of light-curve modulation in some cases, they do not allow the determination of the Blazhko period for any of the variables. Another problem is that the photographic data are seriously affected by daily, monthly and yearly aliases, which may result in an erroneous Blazhko period.

If a modulation period could be determined, its reality was visually checked using animations of the light curves in different phases of the modulation. The animated light curves of the M5 Blazhko variables and the data sets that were used are available at the url: {http\\:www.konkoly.hu/24/publications/M5}.

\section{Blazhko stars in M5}
We identified 20 Blazhko candidates among the 50 RRab variables
that were studied.  The properties of the individual stars
are discussed below.

\begin{figure*}
\includegraphics[angle=0,width=17.8cm]{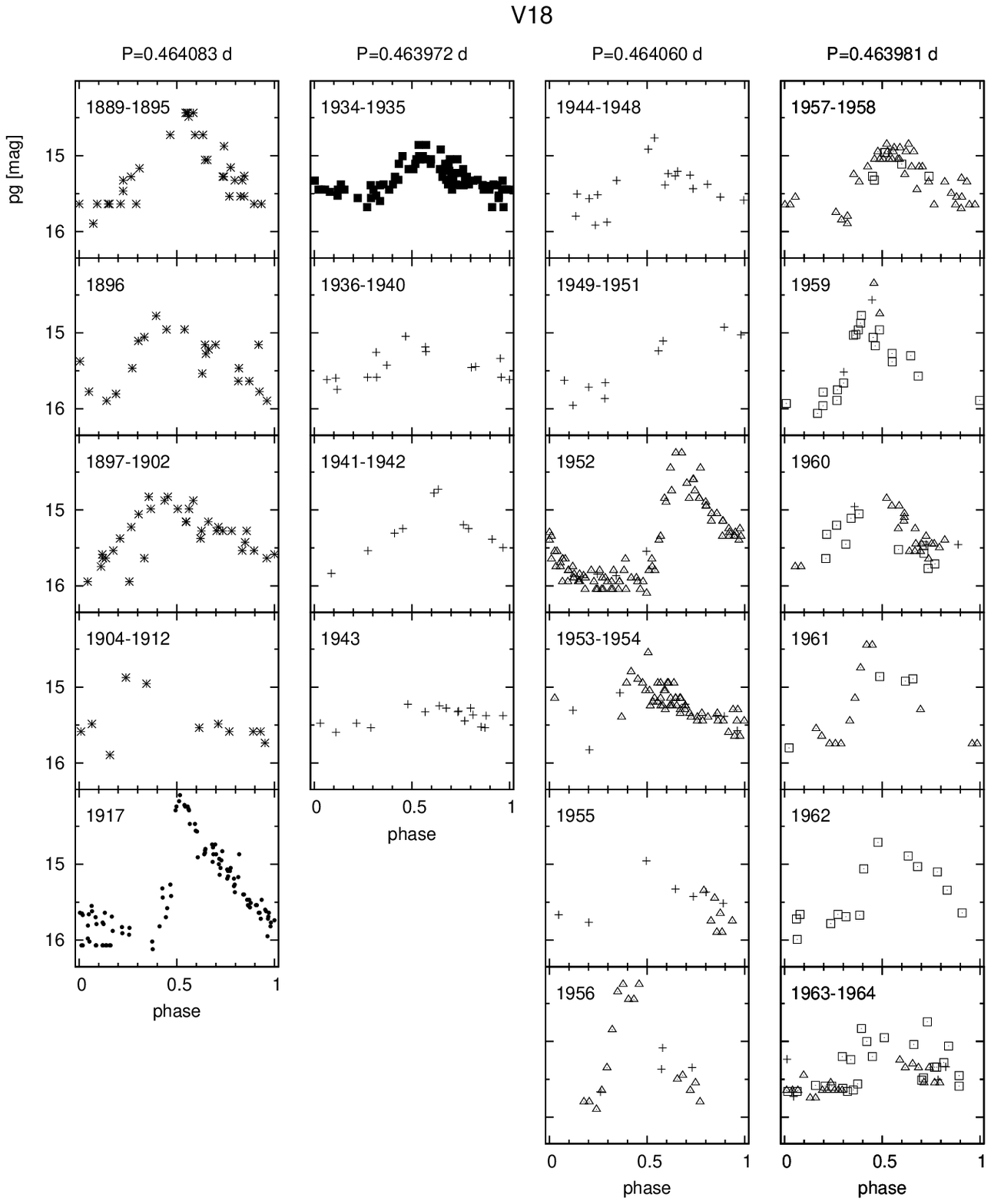}
\caption{Seasonal light curves of V18 between 1889 and 1964. Different symbols denote data from different sources: star \citep{ba17}, dot \citep{sh27}, filled square \citep{oo}, plus \citep{cs69}, triangle \citep{kk}, open square (Konkoly data from Paper I), circle (Las Campanas data from Paper I) and cross  \citep[CCD $B$,][]{c87,s91}. The light curves are phased with the periods given on the top of the columns. The light curves in different years are coherent, their scatter hardly exceeds the observational uncertainty. On the contrary, from one season to the next, the amplitude and the phase of the light curve change significantly; the amplitudes of the amplitude and phase variations are larger than 1-mag and 60-min, respectively.}
\label{v18.1}
\end{figure*}
\begin{figure*}
\includegraphics[angle=0,width=17.8cm]{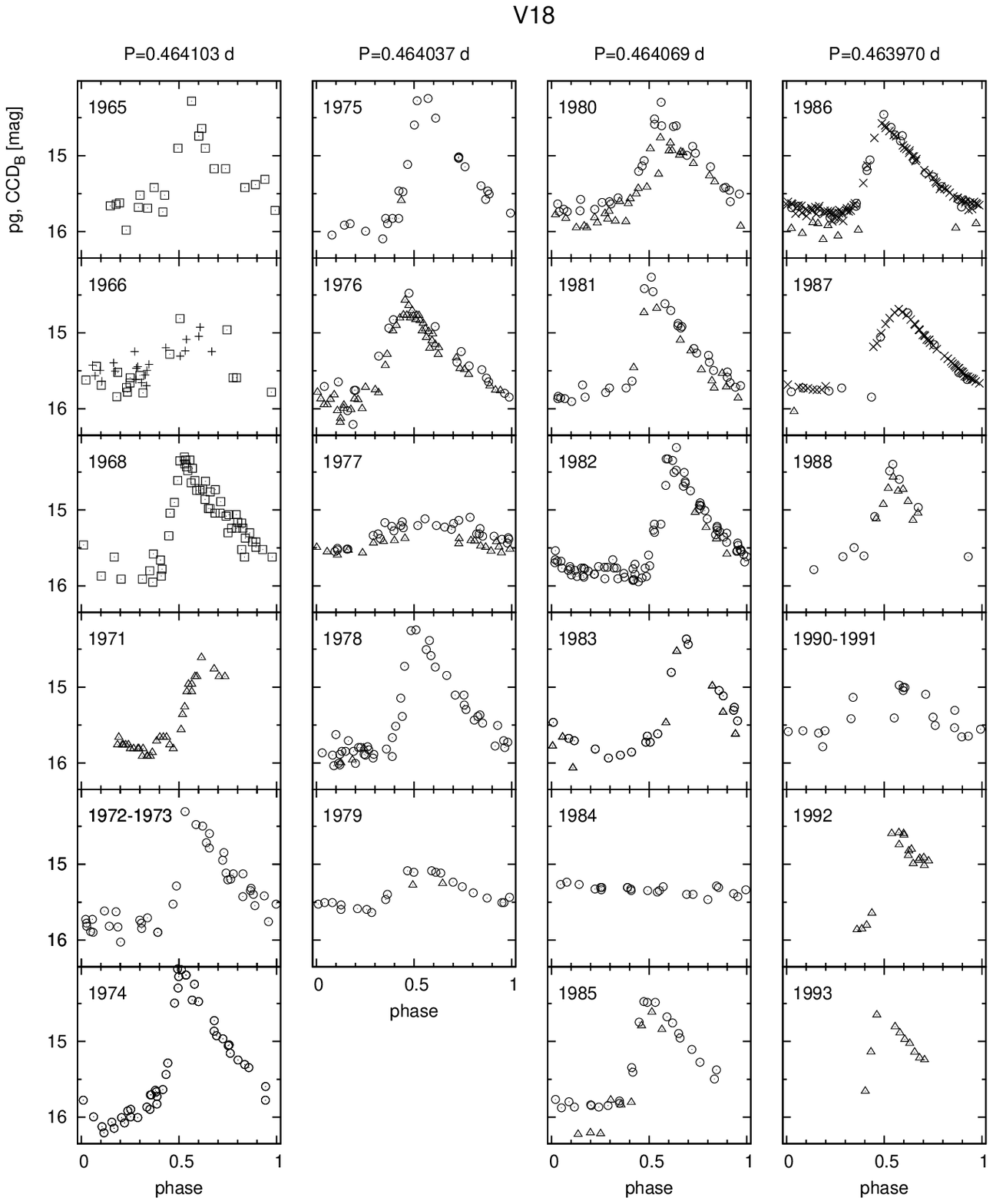}
\contcaption{Light curves of V18 for the 1965--1993 period.}
\end{figure*}
\begin{figure*}
\includegraphics[angle=0,width=17.8cm]{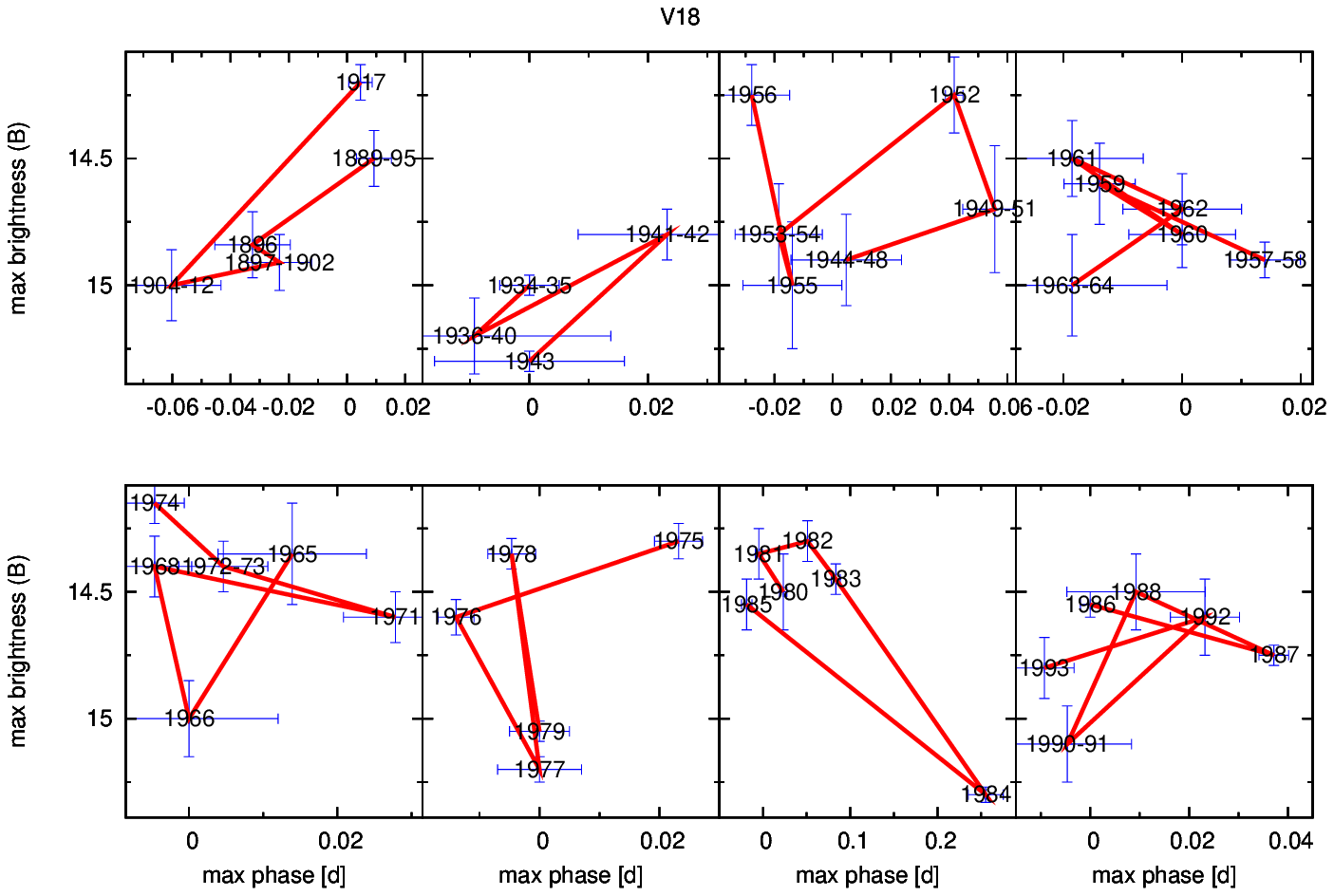}
\caption{Maximum phase and brightness plots are drawn from the seasonal light curves of V18 shown in Fig.~\ref{v18.1}. The periods used to derive the phases of the maxima were those given in the top of the columns in Fig.~\ref{v18.1}. Note the different scales of the phase variations in the different plots. No periodic behaviour of the maximum brightness--maximum phase variations is evident. These plots indicate very complex/irregular light-curve variations on time-scales of some years.}
\label{v18egg}
\end{figure*}

{\bf V1}
There is a slight difference between the shape of minima observed by R96 and K00. The maximum brightness in the R96 data varies between 14.41 and 14.45-mag. The star is far from the centre, so it can be measured accurately on the photographic plates. The residual spectrum of the combined photographic data between JD 2441447 and JD 2449104 (1972--1993) shows a large-amplitude (0.05-mag) signal at $f_0+0.001916$ cd$^{-1}$. We found 522 d as the probable period of the modulation for this time interval. The modulation is dominated by $\approx20$ min phase oscillations. This result is supported by the analysis of the variation of the seasonal $O-C$ values of the entire photometric data (1935--1997) as shown in Fig.~\ref{v01oc}.  Phase oscillations with a period of 512 d fit the observed 0.02-d variation of the seasonal $O-C$ data reasonably well. The CCD data do not contradict this modulation period. The pulsation period of V1 remained constant during the time base of more than 100 years of the observations.

{\bf V2} The star is well separated. The pulsation period did not show any systematic changes during the hundred years of the observations, but irregular period changes of the order of $10^{-5}$ d were detected.
CCD data from K00 show different descending branches. Already Oo41 noted differences between two maxima. A modulation period of 132.38 d was determined by \cite{g76}. We have found clear evidence of the modulation in three segments of the combined photographic data. Between 1971 and 1986, V2 exhibited a very strong modulation with a 136.5-d period. The pulsation amplitude varied between 0.6 and 1.3-mag; the detected changes in the maximum light were 0.6-mag and 75-min in brightness and phase, respectively. The scarce observations from 1966 and 1968 showed a large-amplitude (1.1-mag) light curve, without any evidence of modulation. Between 1956 and 1965 the modulation properties were similar to those found for the 1971--1986 interval, but due to the scarceness of the data, this result is less certain. We also found some indication of the modulation in the data from 1934 to 1955 but with half the period that was found in the other data segments. The pulsation amplitude did not change significantly; it was around 0.8-mag in this part of the data. Any amplitude modulation larger than 0.1-mag was not present, but the amplitude of the phase modulation of the light curve was as large as $\approx$60 min.
 
The residual spectra of V2 in the vicinity of the main pulsation frequency are shown for different segments of the data in Fig.~\ref{v02sp} .

\begin{figure}
\includegraphics[angle=-90,width=8.2cm]{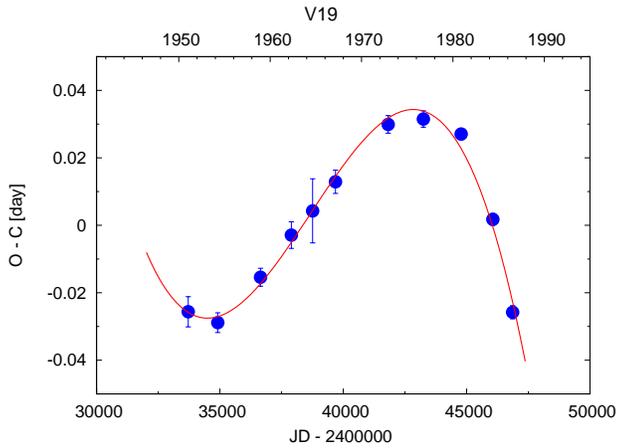}
\caption{Enlarged plot of the $O-C$ variation of V19 for the 1950--1986 period. The data can be accurately fitted with a third-order polynomial as shown in the figure. This polynomial fit is applied to eliminate the period variation from the data when searching for a Blazhko period of the light variation.} 
\label{v19oc}
\end{figure}
\begin{figure}
\includegraphics[angle=0,width=8.4cm]{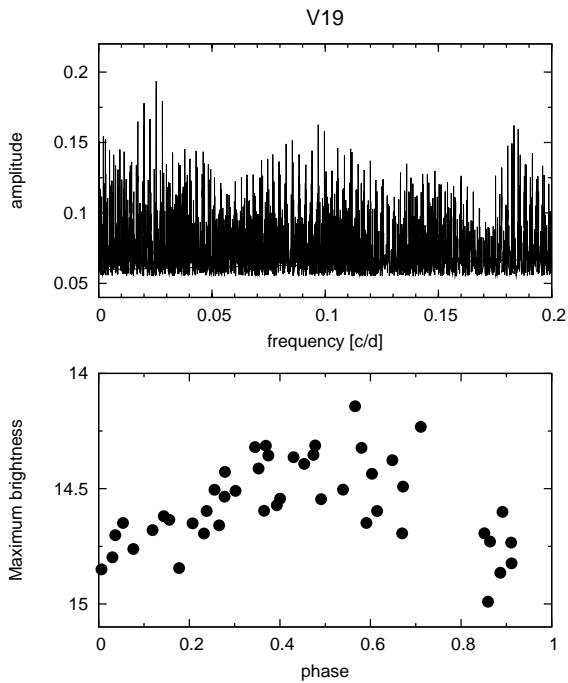}
\caption{The amplitude spectrum of V19's maximum brightness data derived for the 1952--1987 data shows the largest signal at 0.0254 cd$^{-1}$ frequency, corresponding to a 39.4-d periodicity of the amplitude variation. The bottom panel shows the maximum brightness data folded with this modulation period.} 
\label{v19max}
\end{figure}
\begin{figure}
\includegraphics[angle=0,width=7.2cm]{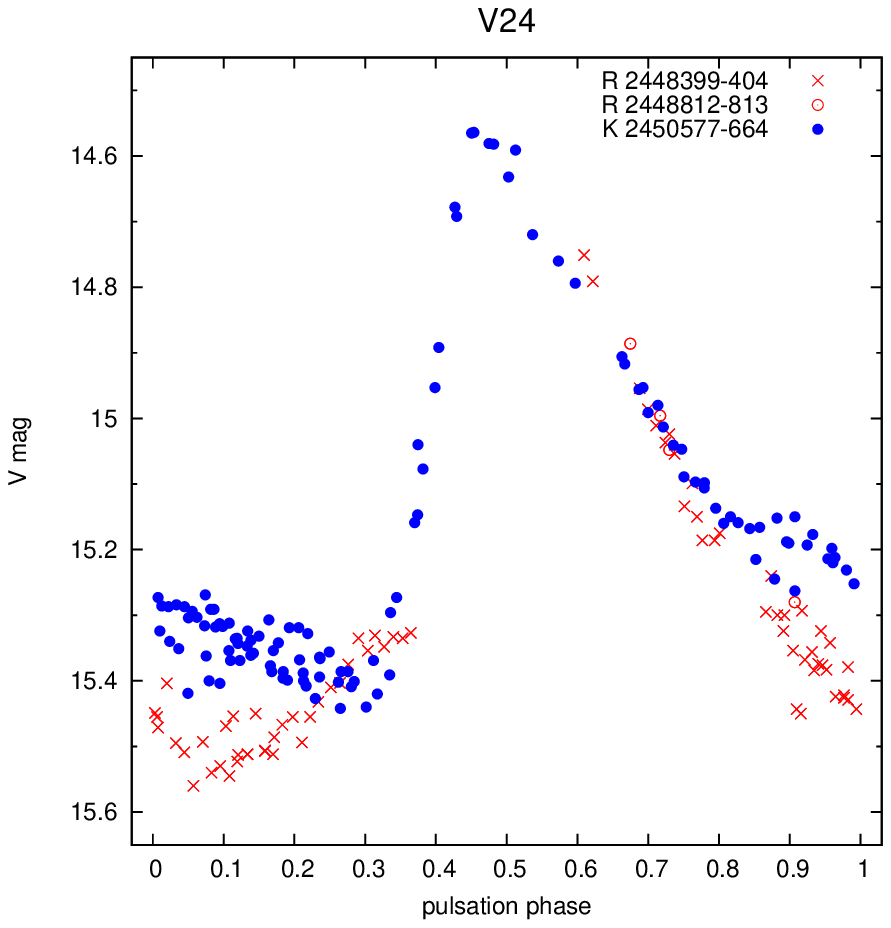}
\caption{CCD $V$ light curve of V24. The different parts of the data are plotted with different symbols as indicated by the key labels. The prefixes R and K refer to the R96 and K00 data, respectively.}
\label{v24}
\end{figure}
\begin{figure}
\includegraphics[angle=0,width=7.2cm]{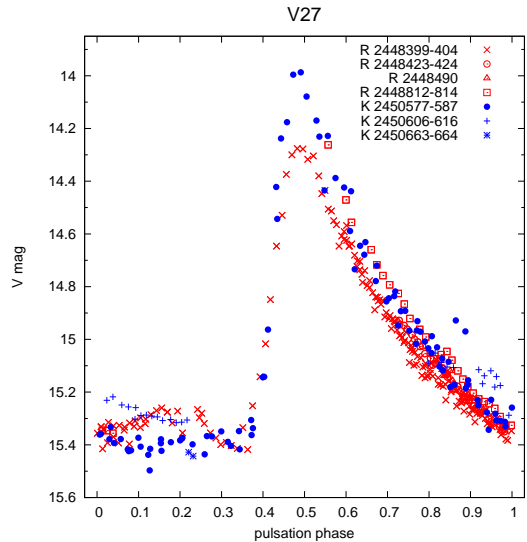}
\caption{CCD $V$ light curve of V27. The different parts of the data are plotted with different symbols as indicated by the key labels. The prefixes R and K refer to the R96 and K00 data, respectively.}
\label{v27}
\end{figure}

{\bf V4} This is the shortest-period RRab star in M5, close to the cluster centre, in a relatively crowded region. Contamination of close companion stars makes the photographic photometry noisy. Oo41 noted RW Dra-type variation of the light curve. The CCD light curve published in R96 indicates phase variation of the rising branch. The pulsation amplitude is larger in R96 than in the K00 data. The pulsation period had been steadily increasing prior to 1970, then a rapid period jump of about 0.00005 d occurred. We searched for a modulation period in the combined photographic data prior to and after the period jump. Between JD 2442553 and JD 2449104 we found clear signals of modulation with a 110-d period (see Fig.~\ref{v04sp}). The light curve was dominantly phase modulated. The light-curve modulation before 1970 was not so evident, however, a 107-d solution was possible. 

{\bf V5} The photographic data are very noisy because of crowding. R96 noted some evidence of light-curve variability. The R96 light curve has somewhat larger amplitude than the K00 one. Based on the combined CCD V data shown in Fig~\ref{v5}, V5 is a possible Balzhko star. The pulsation period was decreasing during the time span of the observations with one of the largest, steady, period-change rates observed in M5.

{\bf V8} The light-curve variability in the phases of the descending branch and at the base of the ascending branch was already noticed by R96.
Differences around minimum light are also indicated by the K00 data.  The amplitude of the CCD $V$ light curves from different observations varies between 0.85 and 1.0-mag.
The star is well separated, therefore these variations are regarded as real. The residual signals appear at similar, 0.0155--0.0150 cd$^{-1}$ separations from $f_0$ in the 1971--1977 and 1978--1993 data, indicating that the modulation period is about 64 d (Fig.~\ref{v08sp}). We failed to detect periodic light-curve modulation in the photographic data before 1971.
The pulsation period has been steadily increasing but the period increase is not linear.

\begin{figure*}
\includegraphics[angle=0,width=17.2cm]{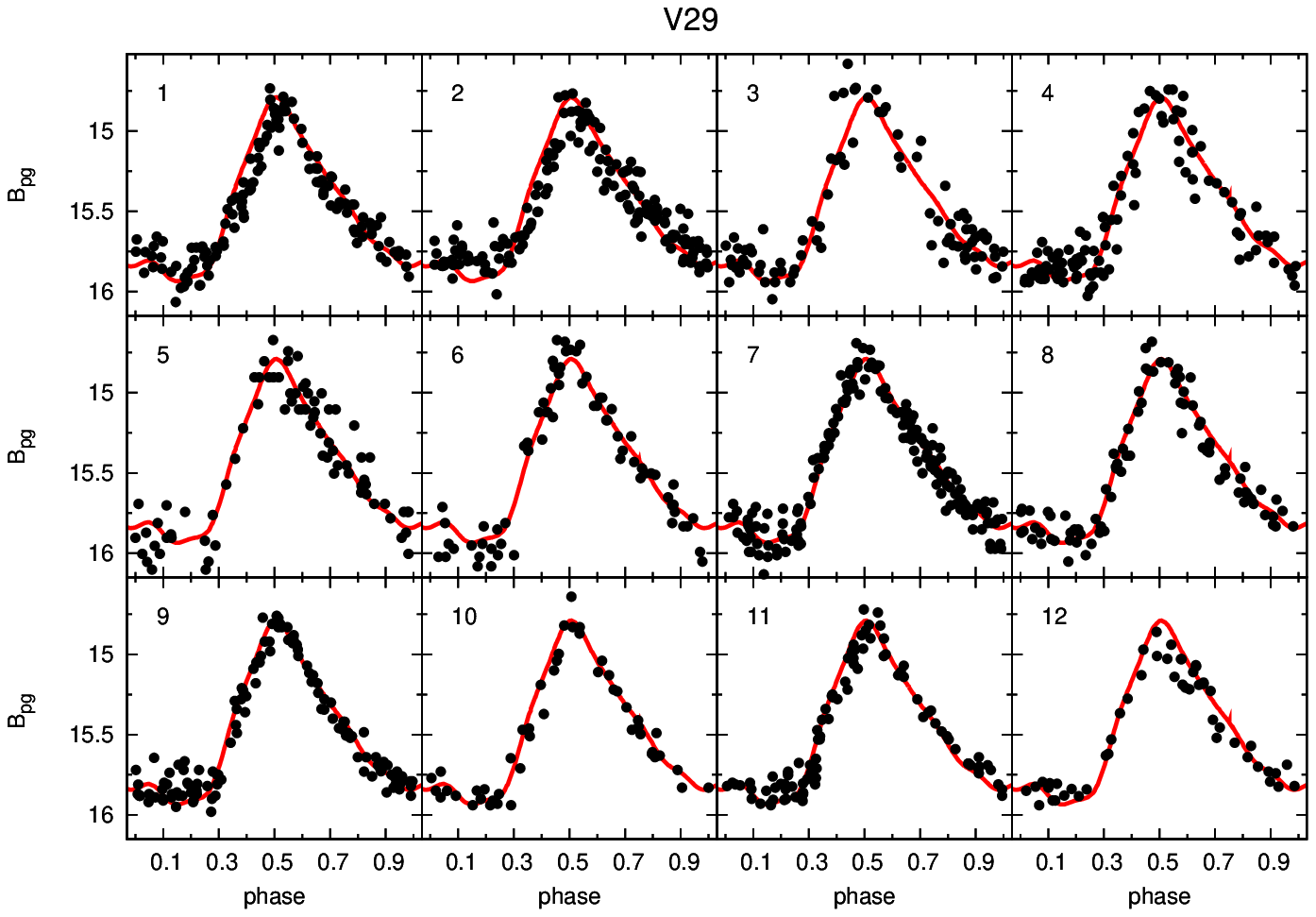}
\caption{The light curves of the 12 subsets of the photographic data of V29. Each subset is phased according to the actual, instantaneous period (see in Fig.~\ref{v29}). The data of panel 7 is fitted with a 7th order Fourier sum, and this sample light curve is drawn for comparison in each plot.  Although some of the light curves are noisy, and no large variation of the light curve is evident, the upper parts of the ascending branch deviate from the sample fit systematically in some of the plots (e.g. in panels 1, 2 and 11).}
\label{v29lc}
\end{figure*}

\begin{figure}
\includegraphics[angle=0,width=8.4cm]{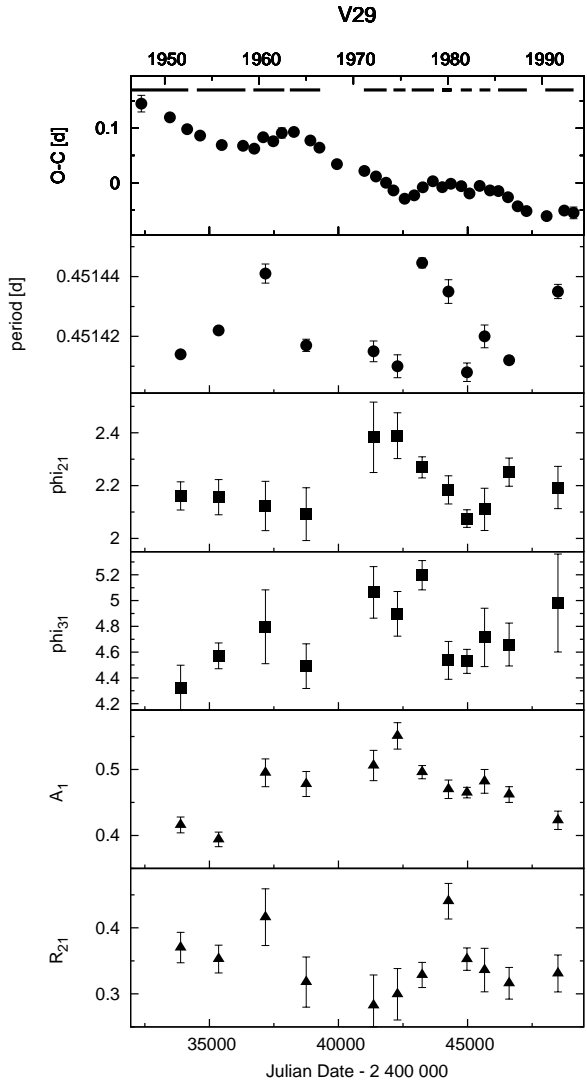}
\caption{The variation of the seasonal $O-C$, the instantaneous periods and the Fourier parameters of 12 (2--5 years long) subsets, shown in Fig.~\ref{v29lc}, of the photographic light curve of V29 are plotted. The time spans of the data subsets are indicated by horizontal lines in the top of the first panel. The $O-C$ shows cyclic fluctuations with about 6000-d period. The consequent changes in the pulsation period are in the 0.00003-d range. The Fourier parameters of the light curves of the data subsets ($\varphi_{21}, \varphi_{31}, A_1, R_{21}=A_2/A_1$) show larger variations than their errors could explain, indicating that the light curve of V29 is not stable.}
\label{v29}
\end{figure}

\begin{figure}
\includegraphics[angle=0,width=8.cm]{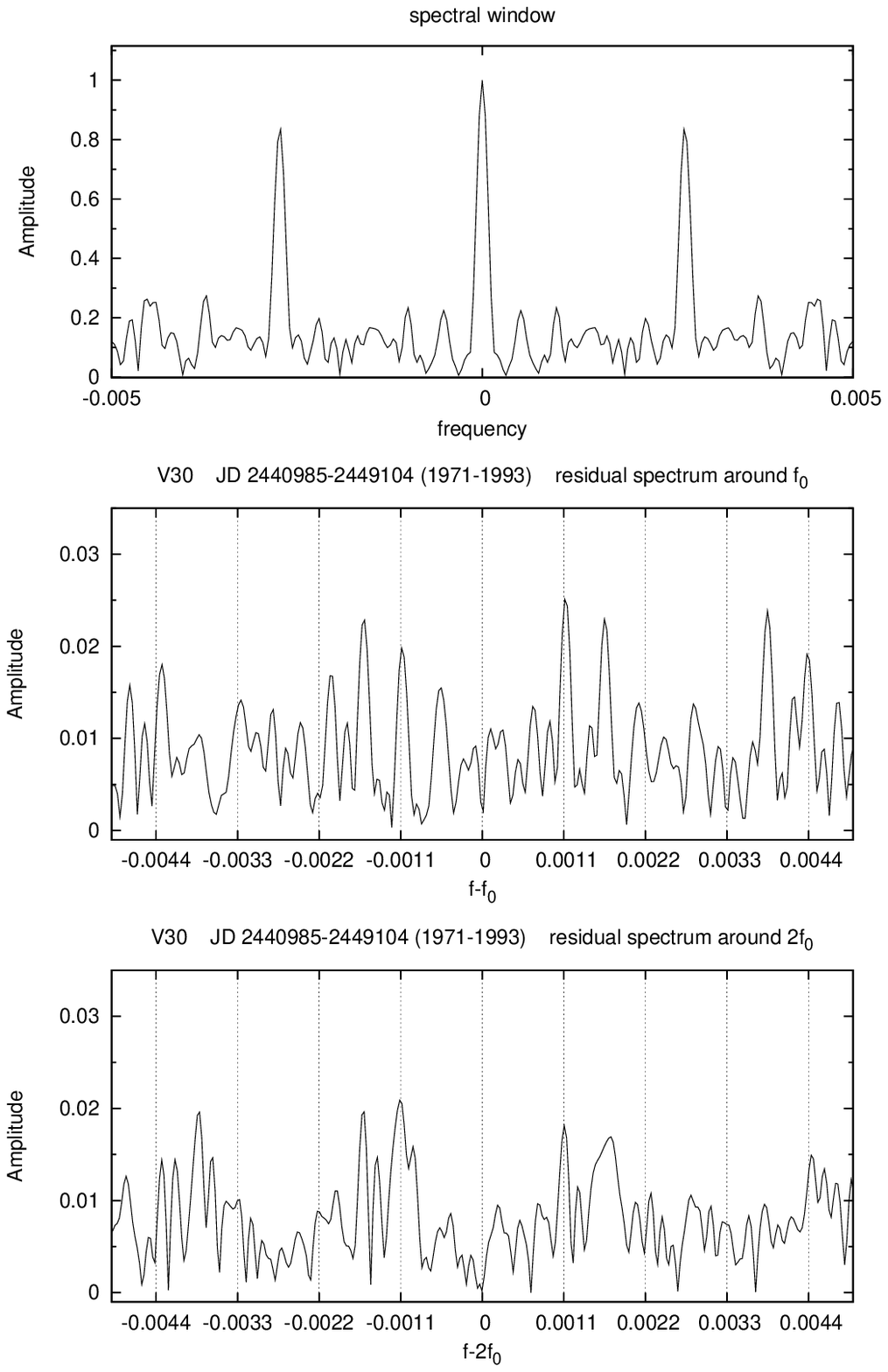}
\caption{The residual spectrum of V30's light curve between JD 2440985 and JD 2449104 are shown. The top panel shows the spectral window. The middle and the bottom panels show the amplitude spectrum of the prewhitened data in the vicinity of $f_0$ and $2f_0$, respectively. Modulation frequencies appear at $f_0-f_{\mathrm{m}}$, $f_0+f_{\mathrm{m}}$ and $2f_0+f_{\mathrm{m}}$, $f_{\mathrm{m}}=0.0011$ cd$^{-1}$, corresponding to an $\approx900$ d modulation period. A yearly alias solution corresponds to a modulation frequency of $f_{\mathrm{m}}=0.0016$ cd$^{-1}$ (621 d). } 
\label{v30sp}
\end{figure}

\begin{figure}
\includegraphics[angle=0,width=7.4cm]{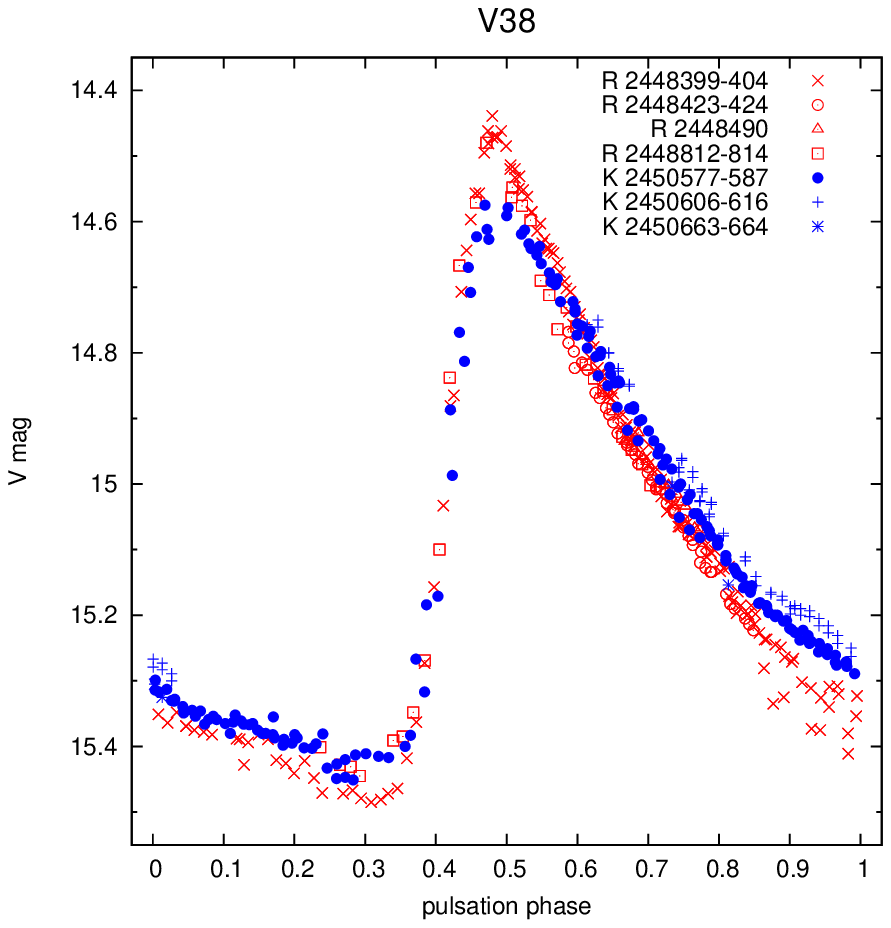}
\caption{CCD $V$ light curve of V38. The different parts of the data are plotted with different symbols as indicated by the key labels. The prefixes R and K refer to the R96 and K00 data, respectively.}
\label{v38}
\end{figure}

\begin{figure}
\includegraphics[angle=0,width=7.4cm]{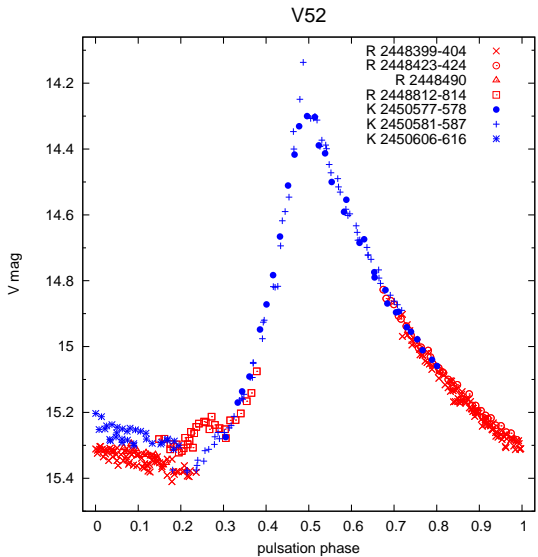}
\caption{CCD $V$ light curve of V52. The different parts of the data are plotted with different symbols as indicated by the key labels. The prefixes R and K refer to the R96 and K00 data, respectively. Light-curve variations at minimum and maximum light are very probable according to the CCD data. } 
\label{v52lc}
\end{figure}

\begin{figure}
\includegraphics[angle=0,width=8.2cm]{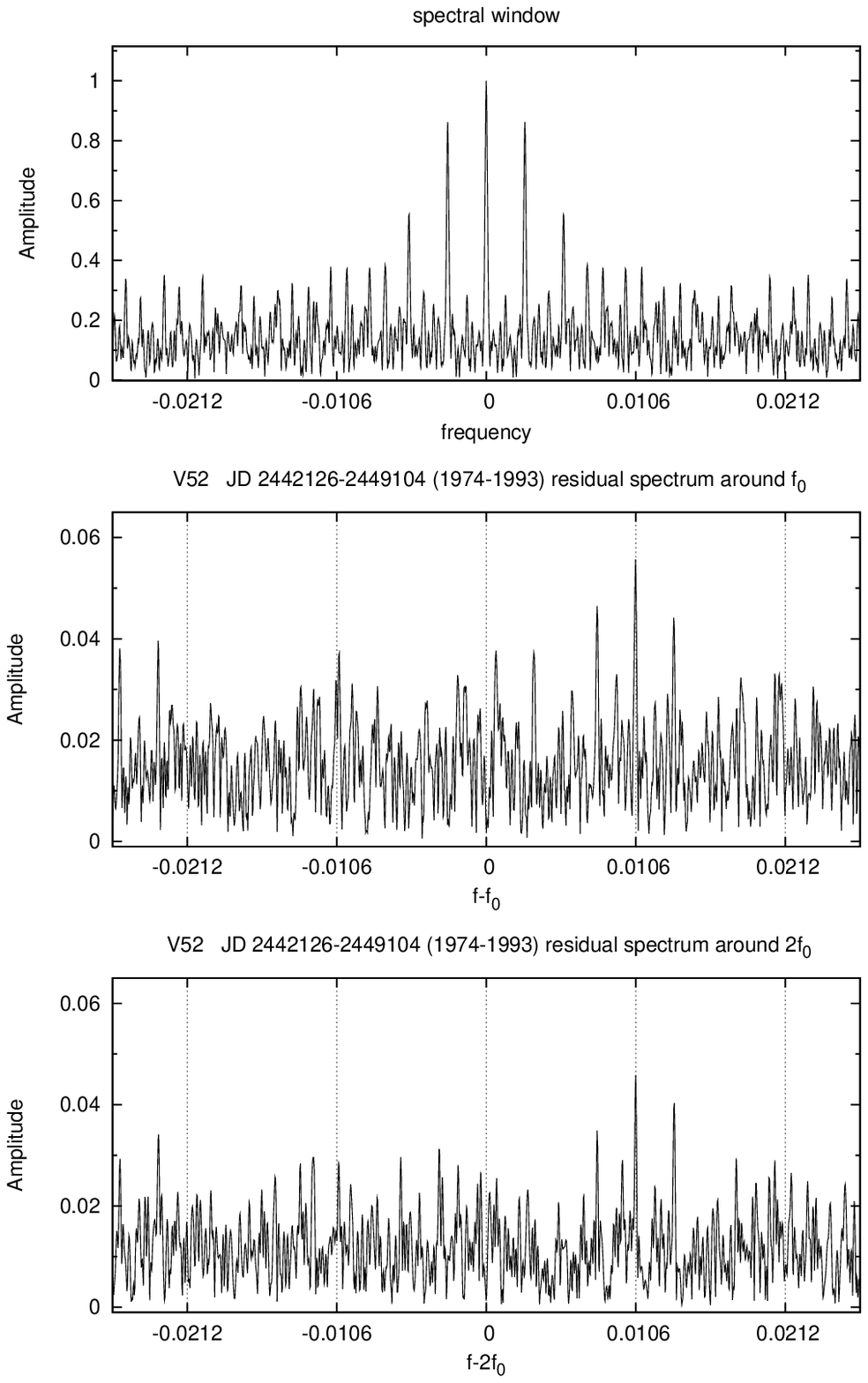}
\caption{The residual spectrum of V52's light curve between JD 2442126 and JD 2449104 are shown. The top panel shows the spectral window. The middle and the bottom panels show the amplitude spectrum of the prewhitened data in the vicinity of $f_0$ and $2f_0$, respectively. Modulation frequencies are apparent at $f_0-f_{\mathrm{m}}$, $f_0+f_{\mathrm{m}}$ and $2f_0+f_{\mathrm{m}}$, $f_{\mathrm{m}}=0.00104$ cd$^{-1}$, corresponding to an $\approx95$ d modulation period.  } 
\label{v52sp}
\end{figure}

\begin{figure}
\includegraphics[angle=0,width=8.2cm]{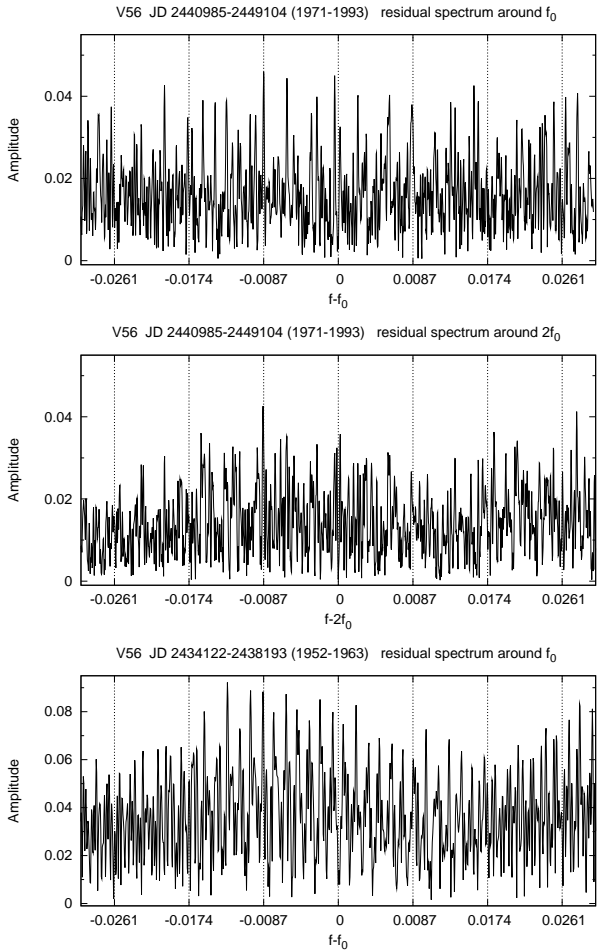}
\caption{The top, the middle and the bottom panels show the residual spectra of V56 in the vicinity of $f_0$ and $2f_0$ for the 1971--1993 period and around $f_0$ for the 1952--1963 period. Although the largest-amplitude signals appear at different frequencies in the three residual spectra, the frequency at $-0.0087$ cd$^{-1}$ separation has significant amplitudes in each spectra. This frequency is identified as $f_{\mathrm{m}}$ (170 d) for V56, however, the one-cycle-per-year alias solution corresponding to $f_{\mathrm{m}}=-0.006$ cd$^{-1}$ cannot be excluded for sure. } 
\label{v56sp}

\end{figure}
\begin{figure}
\includegraphics[angle=0,width=8.3cm]{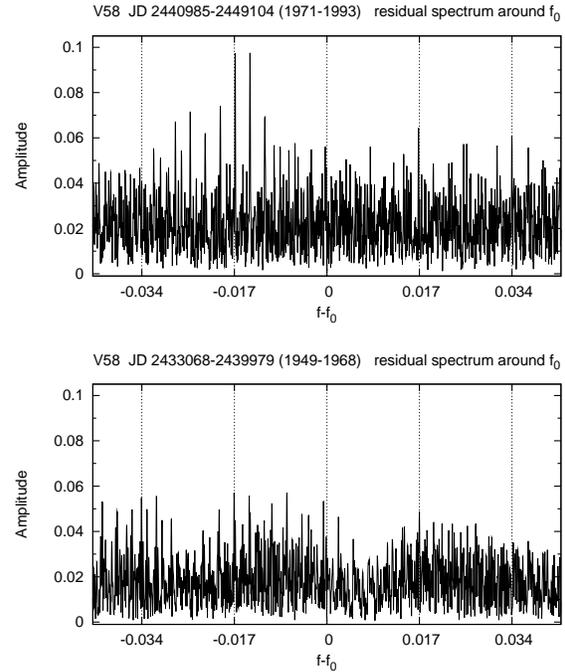}
\caption{The residual spectra around the main pulsation frequency ($f_0$) of the photographic data of V58 for the periods JD 2440985--2449104 (top panel) and JD 2433068--2439979 (bottom panel) are shown. Modulation with a 0.017 cd$^{-1}$ frequency (59 d period) appears in both data sets.} 
\label{v58sp}
\end{figure}

\begin{figure}
\label{v63sp}
\includegraphics[angle=0,width=8.8cm]{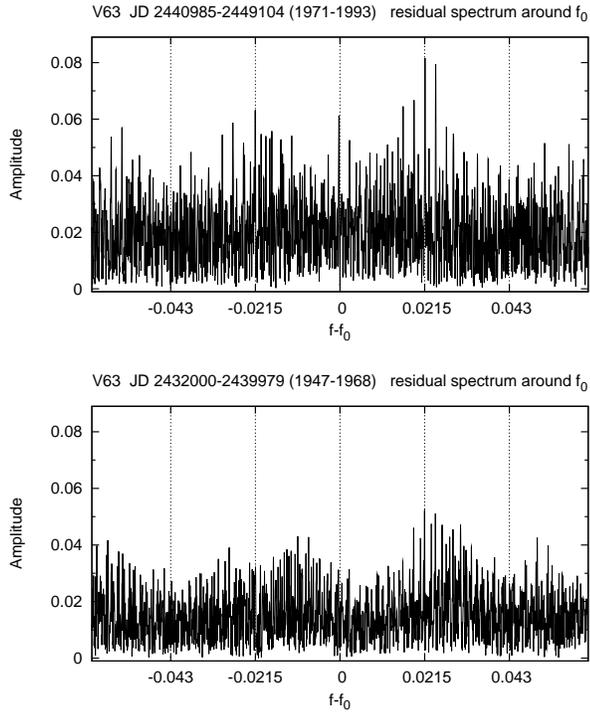}
\caption{The residual spectra around the main pulsation frequency ($f_0$) of photographic data of V63 for the periods JD 2440985--2449104 (top panel) and JD 2432000--2439979 (bottom panel) are shown. Modulation with a 0.0215 cd$^{-1}$ frequency (46 d) appears in both data sets.} 
\label{v63sp}
\end{figure}

\begin{figure}
\includegraphics[angle=0,width=7.4cm]{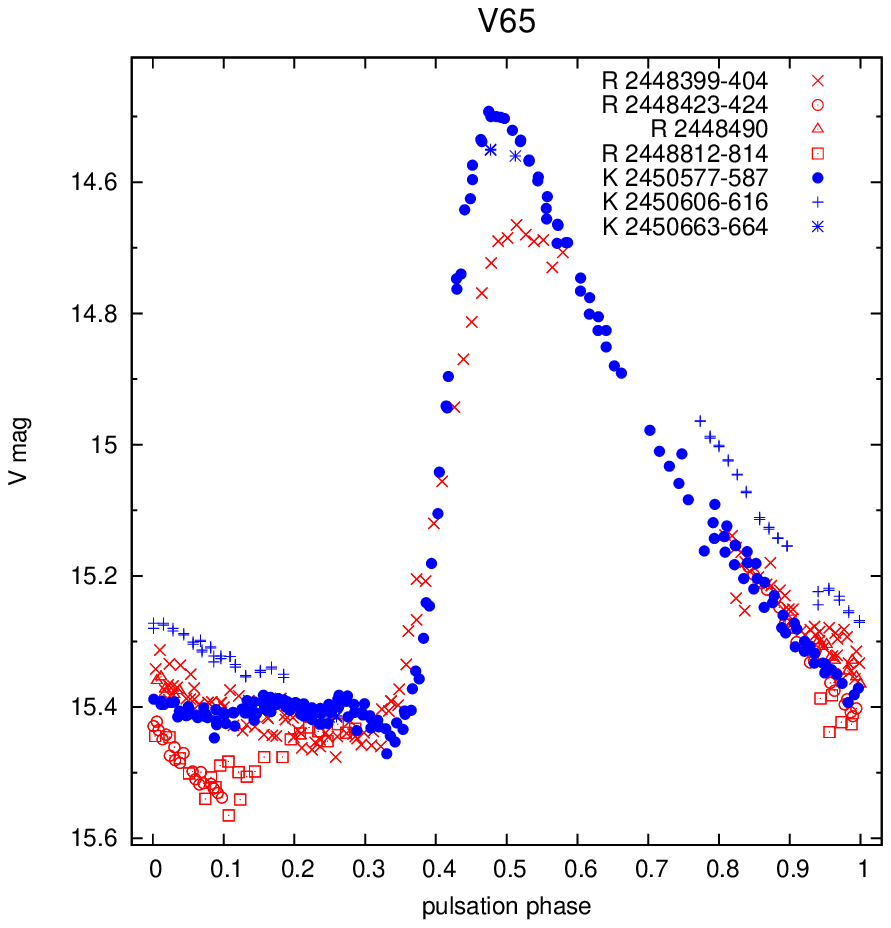}
\caption{CCD $V$ light curve of V65.  The different parts of the data are plotted with different symbols as indicated by the key labels. The prefixes R and K refer to the R96 and K00 data, respectively. }
\label{v65}
\end{figure}

\begin{figure}
\includegraphics[angle=0,width=8.6cm]{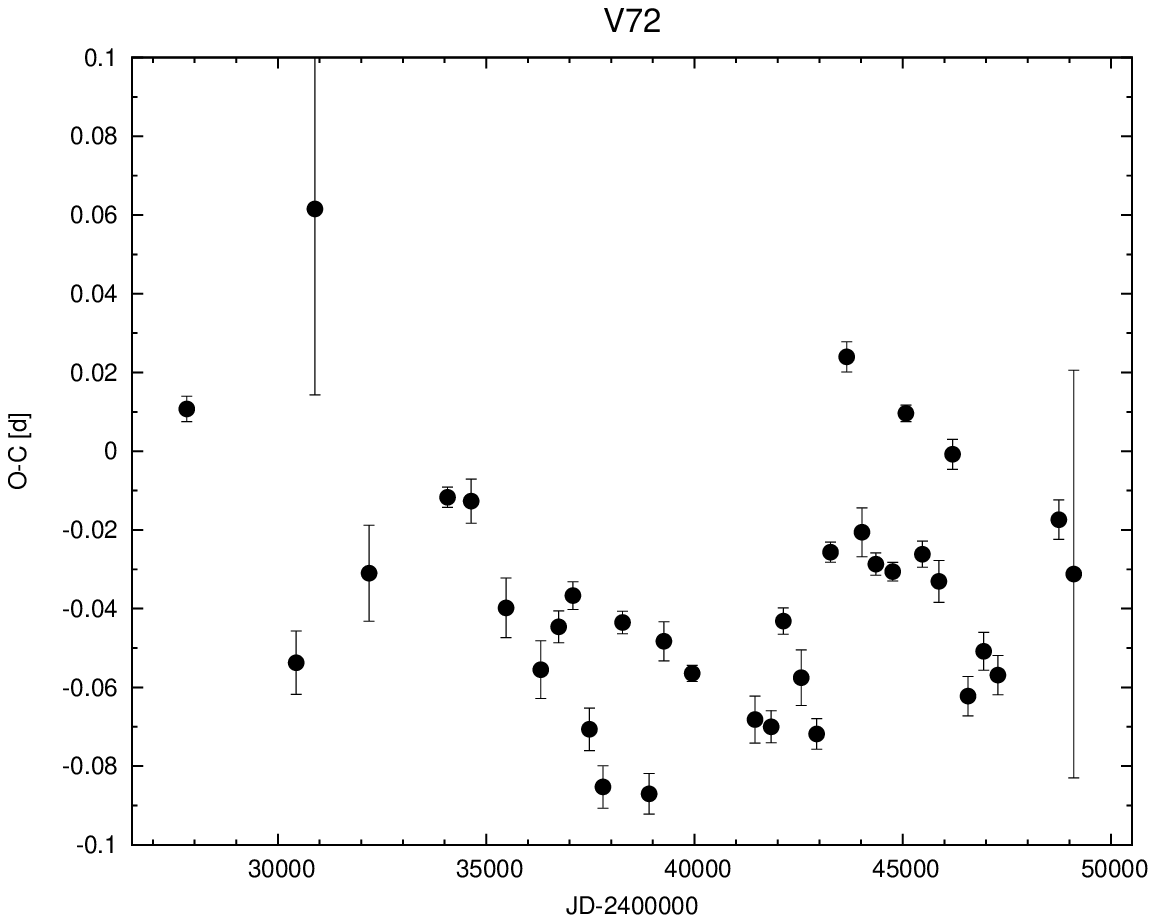}
\caption{The seasonal $O-C$ (phase shift) data of V72 are shown. The $\approx0.05$ d ($>1$ hour) scatter of the data is much larger than the uncertainty of the $O-C$ values. Such rapid, random period variation of an RRab star is unlikely, therefore the significant changes in the phases of the light curve  are most probably caused by Blazhko effect. }
\label{v72oc}
\end{figure}

{\bf V14}
\cite{g80a} determined a 75-d Blazhko-modulation period for V14.
The CCD (R96, K00) light curves show strong amplitude modulation. The maximum-brightness variation is 0.7-mag in the $V$ band. The pulsation-period variation is strong and irregular, which makes the analysis of the light curve difficult. The star can be easily measured; the photographic data are not biased. A 47.8-d (0.021 cd$^{-1}$) modulation period was determined for the  1971--1981 interval (see upper panel in Fig.~\ref{v14sp}). Although strong light-curve modulation was present in the other parts of the photographic data also, a modulation period close to 48 d was not found in any other subset. A 63.8-d modulation period fits the 1949--1959 data reasonably well. The residual spectrum of this data set is shown in the bottom panel of Fig.~\ref{v14sp}. The CCD data do not contradict either the 50 or the 64-d modulation period.

{\bf V18}
The star is far from the centre, but a close companion \citep[IV 75;][]{arp} at 6" distance with brightness similar to V18 at minimum light may contaminate the photometric data of low-resolution observations. 

Oo41 noticed considerable changes in the shape and amplitude of the light curve, which were attributed to probable Blazhko effect in KK71. The pulsation-period variation of the star is very complex as can be seen in Fig.~\ref{v18oc}. The total range of the period variation is larger than 0.0001 d and abrupt period changes occur on time-scales of several thousands of days. The seasonal light curves of all the photographic and $B$ band CCD  data are plotted in Fig.~\ref{v18.1}. The observations from different sources are denoted by different symbols. The pulsation light curve of V18 varies significantly from one season to the next, while the scatter of the light curves in the individual panels does not exceed what is expected from the inhomogeneous, photographic data. The $V$ band CCD  light curves of \cite{s91}, \cite{cm92} and K00 (not plotted in Fig.~\ref{v18.1}) also show significant variations.

The amplitude of the pulsation was smaller than 0.5-mag in 1977 and 1979, while no light variability was detected in 1943 and 1984 at all. There are three stars close to V18: IV74, IV75 and IV76 with 14.60, 15.82 and 17.19 $B$ magnitude, respectively \citep{arp}. The mean magnitude of V18 measured in the 1984 Las Campanas observations is, however, 15.35-mag. This brightness value is close to the mean brightness of V18 in any other season, but it is significantly different from the magnitudes of the neighbouring constant stars. Therefore, we exclude the possibility that wrong identification of V18 can account for the constancy of its light curve in these years. 

As the seasonal light curves of V18 do not show large inhomogeneity, the modulation cycle has to be relatively long, about hundreds of days. However, we failed to determine any modulation period when analysing the light curves of different segments of the data. The phase modulation cannot be separated from the irregular period changes (note the different periods of each column in Fig.~\ref{v18.1}). 
The residual spectrum of any data subset, which is long enough for analysis, is dominated by signals arising from the phase variation and/or the period change. The analysis of the amplitude variation using the maximum brightness data does not give any definite solution for the modulation period, either. The maximum brightness--maximum phase plots of the seasonal light curves also do not help to determine the modulation period (see Fig~\ref{v18egg}). We conclude that the modulation properties and the pulsation period of V18 vary probably on a similar time-scale, making the determination of the modulation period from the available data impossible.

{\bf V19}
Light-curve modulation is probable according to the CCD $V$ data \citep[K00]{s91}. 
The pulsation-period variation is complex, but it can be fitted with a smooth cubic function for the period between 1952 and 1987 as shown in Fig.~\ref{v19oc}. If this part of the combined observations is transformed in time to eliminate the period variation as described in Paper I, the modified data show clear evidence of an amplitude modulation with a $P_{m}\approx39$ d period.
Fig.~\ref{v19max} shows the amplitude spectrum of the maximum brightness data between 1952 and 1987, and the data phased with the largest-amplitude frequency, 0.0254 cd$^{-1}$ (39.4 d). The star is far from the centre with no close companion affecting the photometry.

{\bf V24}
The minima of the R96 and K00 CCD $V$ light curves differ conspicuously  in phase and magnitude as shown in Fig.~\ref{v24}. It clearly indicates that the light curve of V24 is not stable. No maximum light was measured in R96. The star lies in a crowded region, the photographic data are seriously defective. The pulsation-period variation is irregular, because abrupt and continuous period changes both occur.

{\bf V27}
The CCD data shown in Fig.~\ref{v27} confirm the Blazhko effect of V27, which has already been recognized by Oo41. Because of crowding problems and irregular fluctuations of the pulsation period, no conclusive result for the modulation period can be drawn from the photographic data. 

\begin{figure}
\includegraphics[angle=0,width=8.6cm]{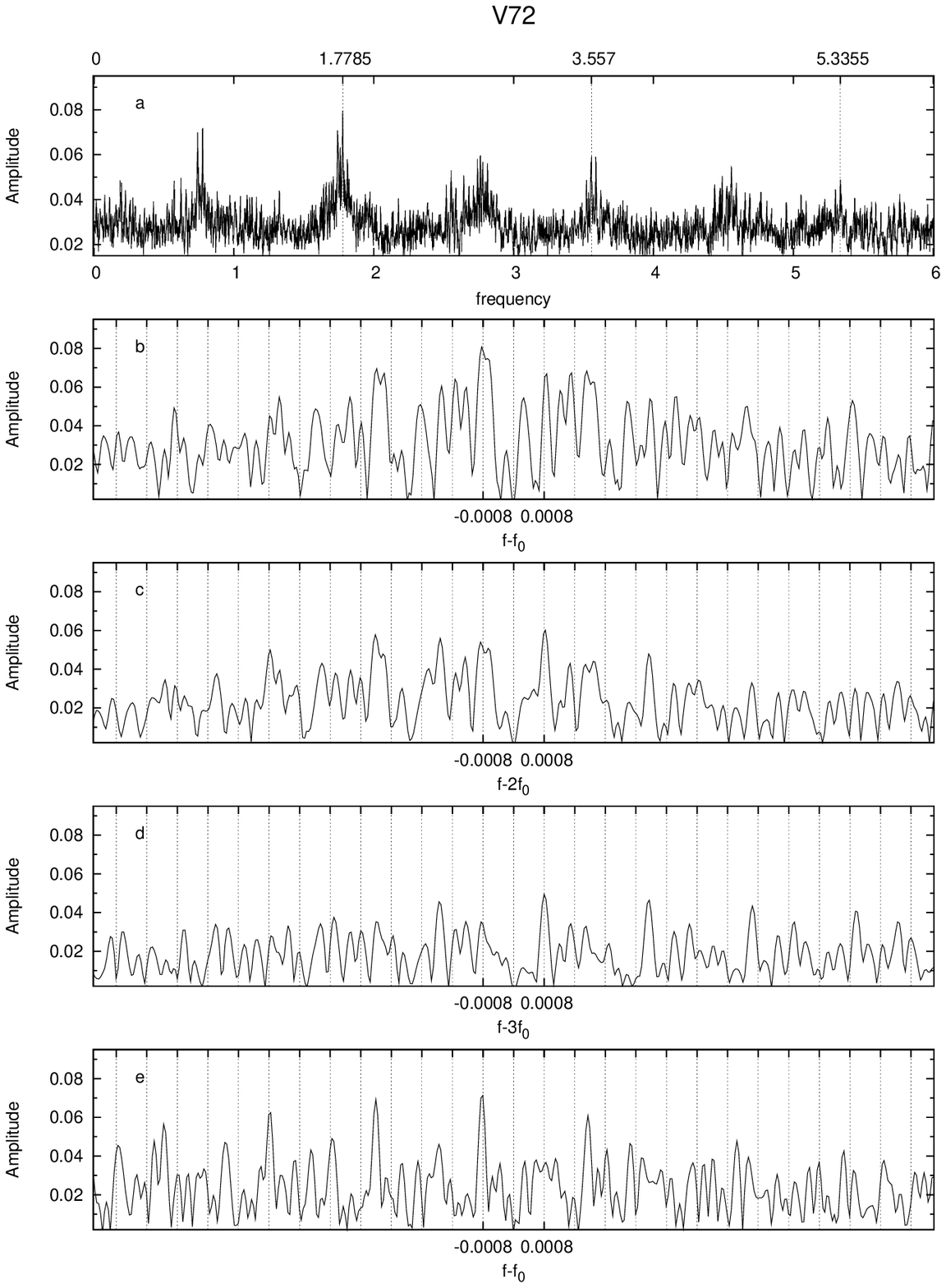}
\caption{The Fourier spectrum of the residual data of V72 for the 1977--1991 period (top panel) is shown. Signals close to the pulsation frequency and its harmonics ($f_0, 2f_0, 3f_0$) dominate the spectrum. The panels $b$, $c$ and $d$ enlarge the residual spectrum at around the pulsation frequencies. The modulation frequencies with 0.0008 cd$^{-1}$ separation, corresponding to a 1200-d period of the modulation, at $f_0-f_{\mathrm{m}}$, $f_0+f_{\mathrm{m}}$, $2f_0-f_{\mathrm{m}}$, $2f_0+f_{\mathrm{m}}$ and $3f_0-f_{\mathrm{m}}$ are the dominant frequencies, although yearly alias solutions with  0.0035 cd$^{-1}$ or 0.0019 cd$^{-1}$ separations are also possible. The bottom panel shows the residual spectrum at $f_0$ of another subset of the data (1955--1968). The signal at $-0.0008$ cd$^{-1}$ or at $-0.0035$ cd$^{-1}$ separation is the largest-amplitude signal of the entire residual spectrum in this part of the data, too.}
\label{v72}
\end{figure}

{\bf V29} A separate star, far from the cluster centre. The CCD light curve (K00) has a somewhat anomalous shape and small amplitude. The anomalous shape of the light curve was already recognised by Oo41.
V29 is the second shortest-period RRab star in M5 ($P_0=0.4514$ d). 
Light curves belonging to the 12 subsets of the photographic observations are shown in  Fig.~\ref{v29lc}.  Variation especially on the upper part of the ascending branch can be suspected.
The pulsation period of V29 decreased by 0.00006 d during the century-long time interval of the observations as shown in Fig.~1 of Paper I. During the second half of the observations, an $\approx7000$ d fluctuation was superimposed on the gradual period decrease.
This is illustrated in the top panel of Fig.~\ref{v29}, which is a
plot of the $O-C$ data from 1946 to 1993. Though no significant light-curve variation is detected, the Fourier parameters of the 2--3 year subsets of the photographic data show  variations  larger than their uncertainties would indicate (Fig.~\ref{v29}). No clear correlation between the variations of the Fourier parameters is evident; however, the changes in the ligh-curve parameters of Blazhko stars can be quite complex in some cases (see e.g. fig. 8 in \citet{ssc} and fig. 12 in \citet{mwI}). Therefore, we suppose that V29 is probably a Blazhko variable, with modulation either with a period very close to one year or on a very long ($\approx7000$ d) time-scale.

{\bf V30}
The CCD $V$ amplitude of K00 is 0.05-mag smaller than that in the \cite{s91} data.
The star is far from the centre. Maximum-brightness variation in the photographic and CCD $B$ \citep{s91} data between 1971 and 1993 is also probable. The Fourier spectrum of this data set indicates a modulation frequency of 0.0011 cd$^{-1}$ (900 d),  however, a yearly alias solution (621 d) is equally  possible (see Fig.~\ref{v30sp}). The pulsation period of V30 was constant during the century-long time base of the observations.

{\bf V38}
The light curves are different in the R96 and K00 data as shown in Fig.~\ref{v38}. Companions at 5"--10" affect some of the photometric data. Because of irregular fluctuations in the pulsation period, we failed to find the period of the modulation from the photographic data. The pulsation period varied irregularly with a $\pm0.00002$ d amplitude.

{\bf V52}
Very faint stars are close to V52, a brighter star is at 12" distance. 
The light curve of V52 is somewhat anomalous, with smaller amplitude and less steep rising branch than that other RR Lyrae stars with similar period have. The pulsation-period variation is complex; abrupt and continuous changes of the period are detected. 
Oo41 classified the star as irregular. The CCD data shown in Fig.~\ref{v52lc} indicate light-curve variability. The Fourier spectrum of the residual photographic data between 1974--1993 show modulation-frequency components at $f_0-f_{\mathrm{m}}$, $f_0+f_{\mathrm{m}}$ and $2f_0+f_{\mathrm{m}}$ with  $f_{\mathrm{m}}=0.00104$ cd$^{-1}$ (95 d) (Fig.~\ref{v52sp}). The modulation is the most prominent in the phase and steepness of the rising branch.

{\bf V56}
The CCD light curves of R96 and K00 have different shapes and amplitudes. The maximum-brightness variation is about 0.2-mag in $V$ band. A very faint close companion does not affect the photometry seriously.
The photographic data show strong maximum-brightness variation, too. Modulation-frequency components corresponding to the same, a 115-d modulation period appear in the residual spectra of both the 1971--1993 and 1952--1963 photographic data, however, a 170-d alias solution cannot be excluded (Fig.~\ref{v56sp}). The CCD data do not contradict the 115-d modulation period. The period changes indicate long-term, cyclic variation.

{\bf V58}
A separate star, far from the cluster centre. No CCD observation is available. \cite{cs69} classified V58 as irregular, while KK71 noted its Blazhko effect. The photographic data indicate modulation with a 59-d period (Fig.~\ref{v58sp}). Strong phase and amplitude modulations are detected. The pulsation period of V58 has monotonously decreased by 0.00002 d during the one-hundred-years of the observations.

{\bf V63} The star is far from the centre, there is no special source of photometric distortion. RW Dra-type variability was recognized by Oo41, and \cite{g80b} determined a 146.8-d Blazhko period of the amplitude variation. The CCD data from K00 cover only the maximum and descending branch of the light curve with $\approx0.1$ mag differences. Modulation with a 46-d period can be detected in two different parts of the photographic observations (Fig.~\ref{v63sp}). Strong amplitude and phase modulations of the light curve are evident in both data sets. The pulsation period has been steadily increasing, but the instantaneous periods show large scatter probably connected to the Blazhko modulation of the star.

{\bf V65}
A close (3") companion with  brightness similar to V65 at minimum light makes the photographic data unsuitable for detecting light-curve variability. The CCD data show light-curve variation both at maximum and minimum light (Fig.~\ref{v65}). The range of the pulsation period variation was about 0.000025 d; continuous and rapid changes were observed.

{\bf V72}
A separate star at large distance from the cluster centre. Oo41 classified the star as irregular. No CCD observation of V72 is available. The pulsation period increased 0.00014 d abruptly sometime between 1917 and 1934, when no observation was obtained. Since 1934, the pulsation period has not changed systematically, but variations of about 0.05 d were detected in the yearly $O-C$ data (Fig.~\ref{v72oc}). The Fourier spectra of the residuals of two different segments of the data (1977--1991 and 1955--1968) have the largest-amplitude signals close to the main pulsation frequencies, which can be attributed to either pulsation-period variation or long-period Blazhko modulation of the light curve. The dominant signals are, however, at the same ($-0.008$ cd$^{-1}$) separation from $f_0$ in both residual spectra as shown in Fig.~\ref{v72}. Therefore, we conclude that some of the scatter in the $O-C$ of V72 is the consequence of light-curve modulation with about a 1200-d period.

{\bf V97}
The CCD light curve (R96) shows larger than 0.3-mag amplitude variation, and indicates that the period of the modulation is longer than 50 d. Photographic data are confined to the Oo41 and the Konkoly 1-m (Paper I) observations. Though crowding affects the photographic data seriously, a modulation period of about 211-d (or its yearly alias corresponding to 499 d) is presumed in the Konkoly data. Due to the sparseness of the observations, no definite conclusion on the pulsation-period variation of the star can be drawn.

\section{Discussion}

\subsection{General properties of the Blazhko variables in M5}

We searched for the Blazhko effect in the collected photometric data of 50 RRab stars in M5. As a result, we found clear evidence of the Blazhko effect in 18 cases, and, with less certainty, in two additional cases. However, small-amplitude modulation cannot be excluded in the remaining 30 RRab stars either, especially in the ones that lack CCD observations. Therefore, the detected 40 per cent incidence rate is a lower limit of the true occurrence of Blazhko stars in M5. This is in good agreement with the 
results of the recent, high-accuracy surveys of RRab stars: 47 and 40 per cent incidence rates of modulated light curves have been found in a sample of short-period ($P< 0.5$ d) variables \citep{stat} and in the {\it Kepler} field \citep{kepler,benko}, respectively. 

The modulation periods could be determined for 13 stars. The results are summarized in Table~\ref{bltab}. The variable, the time interval of the data used to identify the modulation, the pulsation and the modulation periods, the amplitudes of the amplitude and the phase modulations, the intensity-averaged mean $V$ brightness (see Paper I) and the type of the pulsation-period change are given in the columns.

The formal errors of the modulation periods are 0.1--0.2 and 1--2 d for variables with Blazhko periods shorther and longer than about 100 d, respectively. However, the incompleteness of the data sampling resulted in alias solutions which were equally acceptable in some cases. The alias solutions corresponded to modulation periods that differed by tens/hundreds of days. If the selection of the modulation period was somewhat ambiguous, this was noted in the previous section, and  the other possible modulation periods were also given.

The modulation properties are simply read from the folded light curves. Thus they give only rough estimates of the amplitudes of the amplitude and phase modulations, which might, in fact, be even larger for some of the Blazhko stars.

The mean $V$ magnitudes are derived from the CCD data of a narrow 
modulation-phase bin. They are estimated to be accurate within about 0.01--0.03 mag. The uncertainties of the mean $V$ magnitudes arise from zero-point errors/differences in the CCD data, from the incomplete phase coverage and from the possible changes in the mean brightness in the different phases of the Blazhko cycle ($<0.02$ mag).

The period distribution of the Blazhko variables is different from the distribution of the total sample of RRab stars in M5. The Blazhko effect favours shorter periods. The mean period of all of the RRab stars is 0.546 d, but for the Blazhko variables it is only 0.504 d. 
In the sample of RRab variables with periods shorter than 0.55 d 
and photometric data suitable for detecting light-curve variability,
the occurrence rate is as high as 60 per cent. This is not a unique property of M5; \cite{pr} already recognized that `there may be a tendency for such variables to occur at slightly shorter periods than the class as a whole.'

\begin{figure*}
\includegraphics[angle=0,width=16.6cm]{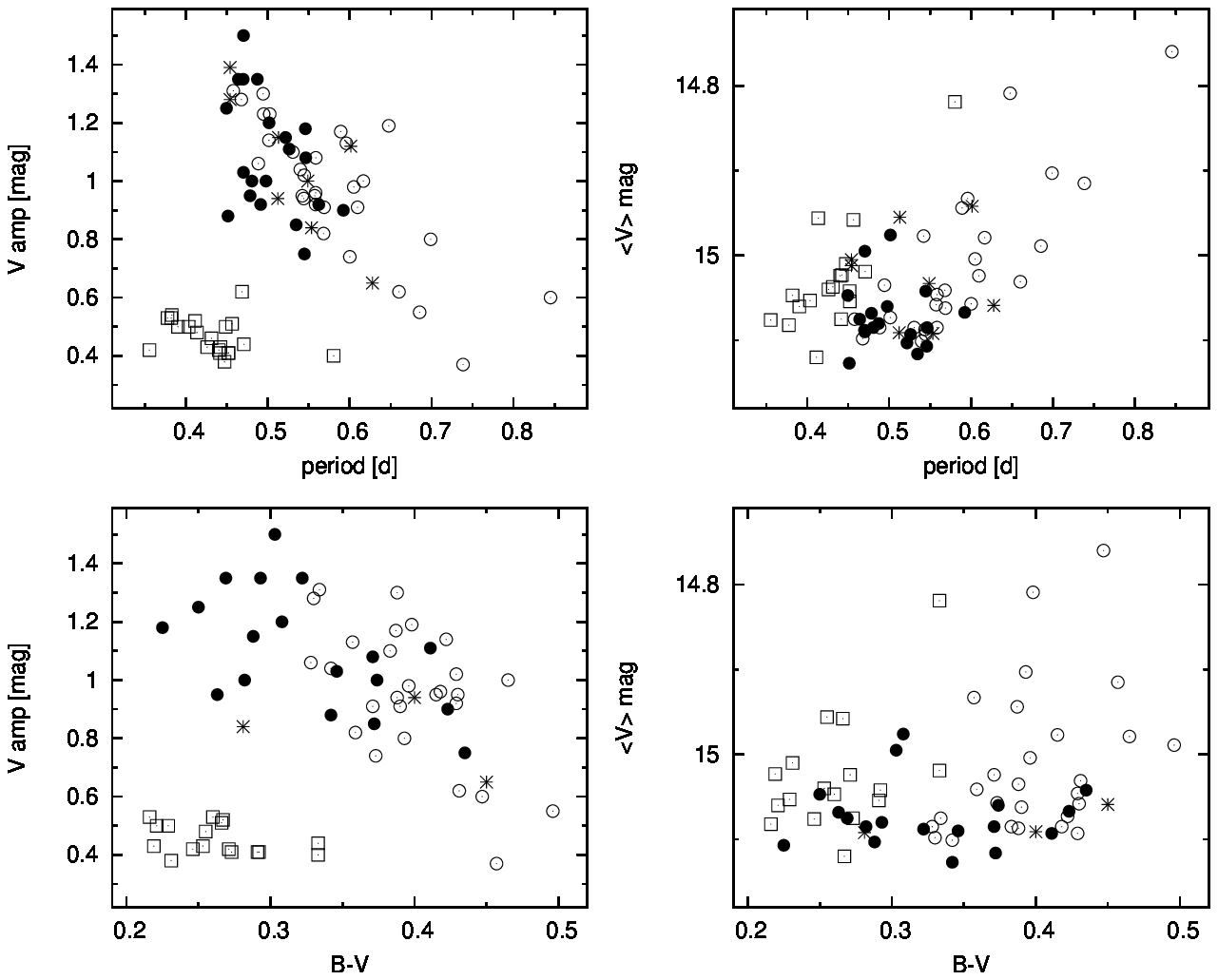}
\caption{The amplitude and mean $V$ brightness distributions of RR Lyr stars in M5 versus period (top panels) and versus $B-V$ colour index (bottom panels) are plotted. Filled and open circles denote the Blazhko and the regular RRab stars. The variables with photometric data inappropriate to detect light-curve modulation are shown by asterisks. First-overtone RRc stars are shown by squares, their periods are fundamentalized. The RR Lyrae stars with Blazhko modulation tend to have systematically larger amplitude, shorter period and fainter magnitude than the averages of the whole sample.
The $B-V$ colour indices of the Blazhko stars overlap with that of the RRc variables.  }
\label{sum}
\end{figure*}

In Fig.~\ref{sum} the mean $V$ magnitudes and the $V$ amplitudes of M5 RR Lyrae stars are plotted against pulsation period and the $B-V$ colour index. Data are taken from Table 4 in Paper I. The  observed largest amplitudes of Blazhko stars are plotted. The periods of the first-overtone variables are fundamentalized for direct comparison. The Blazhko stars occupy the short-period range of RRab stars, and are adjacent to the first-overtone RRc variables. Their $B-V$ colours overlap with those of the  RRc stars. In fact, among the eleven bluest RRab stars with mean $B-V$ colour indices falling in the regime occupied by the RRc stars, all but one (V83) are identified as Blazhko variables. 
The poor quality of the photometry of the one, exceptional star (denoted by asterisk in Fig.~\ref{sum}), does not allow detection of any light-curve variability of this star.

\begin{table*}
\caption{}
 \label{bltab}
  \begin{tabular}{llccccccc}
\hline
\multicolumn{2}{l}{star}& time interval [JD]&\multicolumn{2}{c}{period [d]}& \multicolumn{2}{c}{ modulation property}&$<V>_{\rm {int}}$&period change\\ 
&&&pulsation&modulation&amp(B) [mag] & phase [min]&&\\
\hline
V1 & & 2441447 -- 2449104 & 0.521787 & 521.9 & 0.1   & 15   &15.103& stable\\
V2 &a& 2440985 -- 2446592 & 0.526260 & 136.5 & 0.6   & 75   &15.093& irreg.\\
   &b& 2435629 -- 2438968 & 0.526277 & 129.7 & 0.6   & 75   &      & \\
   &c& 2427540 -- 2435310 & 0.526262 &  66.7 &$<$0.1 & 60   &      & \\
V4 &a& 2442598 -- 2449104 & 0.449618 & 110.1 & 0.1   & 45   &15.047& irreg.\\
   &b& 2434122 -- 2438193 & 0.449671 & 106.9:& 0.3:  & 30:  &      &  \\
(V5:)& & 2448399 -- 2450664 & 0.545851 &       &$<$0.1(V)&  - &15.107&decreasing\\
V8 &a& 2443631 -- 2449104 & 0.546243 & 64.5: &$<$0.1 & 15   &15.085&irreg.$^{\diamond}$\\
   &b& 2440985 -- 2443286 & 0.546237 & 66.5: & 0.2   & -    &      &\\
V14&a& 2440985 -- 2444767 & 0.487188 &  47.8 & 0.5  & 50   &15.080& irreg.\\
   &b& 2433068 -- 2436762 & 0.487159 &  63.8 & 0.5  & 50   &      & \\
V18& & 2411148 -- 2444767 & 0.464060 &$>$500 & 1.5  & 60   &15.075&irreg.\\
V19& & 2434122 -- 2447298 & 0.469987 &  39.4 & 0.4  & -    &15.088& irreg.\\
V24& & 2448399 -- 2450664 & 0.478445 &       & 0.1$^{*}$(V)& 100$^{*}$&15.068& irreg.\\
V27& & 2431259 -- 2449104 & 0.470312 &       & 0.3(V)& -   &14.995& irreg.\\
(V29:)&& 2448399 -- 2450664 & 0.451427 &       &0.1:   &     &15.127& decreasing$^{\dag}$\\
V30& & 2440985 -- 2448745 & 0.592176 & 901/621& 0.2  & -   &15.067 & stable\\
V38& & 2440985 -- 2444767 & 0.470419 &       & 0.15(V)&    &15.090 & irreg.\\
V52&&  2442126 -- 2449104 & 0.501518 &   94.5& 0.2:  & 45 & 14.976&irreg.\\
V56&a& 2440985 -- 2449104 & 0.534695 &  115.1& 0.5  & 30 & 15.116&irreg.\\
   &b& 2434122 -- 2438193 & 0.534690 &  114.2& 0.6  & 45 & &\\
V58&a& 2440985 -- 2449104 & 0.491249 &  59.3 & 0.4  & 30 & -- &decreasing\\
   &b& 2433068 -- 2439979 & 0.491254 &  58.9 & 0.4  & 30 & &\\
V63&a& 2440985 -- 2449104 & 0.497684 &  46.4 & 0.4  & 30 &15.06: &increasing$^{\dag}$\\
   &b& 2432000 -- 2439979 & 0.497680 &  46.6 & 0.3  & 30 & &\\
V65& & 2448399 -- 2450664 & 0.480671 &       & 0.2$^{*}$(V)& &15.085& irreg.\\
V72&a& 2443225 -- 2448394 & 0.562276 & 1199.7& 0.1 &80&-- &irreg.\\
   &b& 2435251 -- 2439979 & 0.562270 & 1230.8& 0.1 &80&&\\
V97& & 2442598 -- 2449104 & 0.544623 & 211:/499:  &$>$0.4& $>$60&15.042& insuff. data\\
\hline
\multicolumn{7}{l}{\footnotesize{$^{\diamond}$the period is increasing, but the period change is not linear}}\\
\multicolumn{7}{l}{\footnotesize{$^{\dag}$irregular fluctuations superimposed}}\\
\multicolumn{7}{l}{\footnotesize{$^{*}$variation in minimum light}}
\end{tabular}
\end{table*}

Another interesting feature of the location of the Blazhko stars is that they tend to have faint mean-brightness values. The mean values of the brightness of the Blazhko and the regular RRab stars are $15.072\pm0.039$ and $15.015\pm0.081$ mag, respectively. No similar result has been previously shown. However, in most studies of globular-cluster Blazhko stars, the mean brightness has been determined without any distinction of the pulsation/modulation phase coverage of the data. These mean magnitudes may be seriously affected by the incomplete light-curve coverage that is typical in photometric observations of globular-cluster variables. In our previous studies of the field Blazhko variables \citep[e.g. in][]{mwI} it was shown that the mean brightness  determined using a complete pulsation light curve at only one Blazhko phase does not differ more than $\approx0.02$ mag from the true mean brightness. The mean-brightness values of Blazhko variables are determined in this way here, so they can be regarded as relatively accurate.

These specific features of the Blazhko stars raise the possibility that the onset of the Blazhko phenomenon is induced by evolution. One may speculate that the pulsation of RRab stars evolving blueward becomes unstable against the Blazhko effect (whatever it is)  just preceding the mode switch to first-overtone pulsation. A possible relation between the Blazhko and the double-mode stars has been already proposed by \cite{n85}. However, detailed modelling of the mode switch from fundamental to overtone mode is still lacking.

Reviewing the variable populations of the globular clusters rich in RR Lyr stars, one finds that they are either rich in Blazhko stars (NGC362, NGC 3201, M5, $\omega$ Cen) or in double-mode variables (IC4499, M15) or in both (M3, M68).  Blazhko phenomenon is quite common among the galactic-field RRab stars \citep{stat}, but double-mode pulsators are rare. The occurrence of double mode and/or Blazhko stars in the different stellar populations, however, does not show clear correlation with any of their properties e.g. metallicity, HB-type, Oosterhoff-type. 
While Blazhko stars appear in all the globular clusters with large RR Lyrae population, variables showing fundamental-mode and first-overtone pulsation simultaneously are known basically only in M3, M15, M68, IC4499.
Searching for the properties that foster the appearance of double mode or Blazhko modulation would be an important issue in disclosing any possible connection between these phenomena. There is even a unique variable, V79 \citep{v79} in M3, which shares the properties of both the double-mode and the Blazhko phenomena. V79 was a double-mode pulsator for 15 years, between 1992 and 2007. However, observations made in 2008--2009 showed it to be pulsating in the fundamental mode with a 65-d modulation period. Thus this is a good example of a star that confirms Nemec's (1985) proposal.

In Paper I it was already shown that variables with irregular
changes in their pulsation period all show the Blazhko effect,
provided they have photometric data suitable for detecting light-curve
variability. From the 20 Blazhko variables listed in Table~\ref{bltab}, two do not show any period change, one has an increasing period with irregular fluctuations superimposed, one has a period that increases, but not at a uniform rate, two have periods that decrease steadily, one has strong irregularity superimposed on its gradual period decrease and 12 show nothing but irregular period fluctuations. (The remaining star does not have enough observations for characterizing its period-change behaviour.)
Clearly, it can be concluded that the occurrence of the Blazhko modulation is usually accompanied by irregular fluctuations of the pulsation period and vice versa. We note that \cite{st80} proposed a model that could account for both the irregular period changes and the Blazhko effect. He attributed them to hydromagnetic effects in the atmosphere of the pulsating star.

\subsection{Special properties of some of the Blazhko variables in M5}

The amplitude of the modulation of V18 is so large that sometimes it  suppresses the pulsation totally. A similar example is V442 Her \citep{sl} with pulsation amplitude varying between 1.4 and 0.2-mag in the $V$ band. Modulation periods could not be derived for either star. In both stars, significant changes of the light-curve shape and amplitude occur from one season to the next, while the seasonal light curves, spanning $\approx100$ days, are surprisingly homogeneous. The pulsation-period variations are extremely large; random period changes of the order of $10^{-4}$ d are detected in both stars. 
Another globular-cluster variable that exhibits light-curve changes
from one season to the next is the short-period RRab star, V15, in the
Oosterhoff-type I cluster M4 \citep{c94}.
It seems that in these special cases the Blazhko effect manifests itself in strong, random, abrupt changes of the light curves, lacking any clear periodic behaviour. This type of the modulation definitely cannot be explained by the appearence of  non-radial-mode oscillations with frequencies close to the pulsation frequencies.

Other Blazhko stars with a special character are V29 and V72. These stars show hardly any amplitude variation, but their pulsation phase vary considerably on time-scales of some hundreds to thousands of days. The galactic-field analogue of these stars is RS Boo \citep{rsboo}, with $\approx0.15$ mag amplitude and $\approx45$ min phase modulations during its 530-d modulation cycle.
The long-term pulsation-period variation of each of these stars is relatively simple: it is steadily decreasing for V29, fits two constant periods differing by $1.4\times10^{-4}$ d for V72, and has increased throughout the 110 years time base of the observations\footnote{http://rr-lyr.ast.obs-mip.fr} for RS Boo. For V29 and V72 some fluctuations are superimposed on the global trend of their period variations (see also Paper I). The modulations of these Blazhko stars are dominated by cyclic or periodic phase variations. An important issue would be to determine why in some Blazhko stars the modulation is dominated by amplitude variations, while in others it is dominated by phase oscillations.

Significantly different frequencies of the modulation have been derived for different epochs for two Blazhko stars (V2 and V14), which cannot be explained by continuous changes of the modulation period. 
The ratios of the detected modulation periods are close to 2:1 and 4:3 for V2 and V14, respectively. 
Another case of strong multiperiodicity of the modulation has been documented in \citet[][CZ Lacertae]{czl}. The modulation frequencies detected in CZ Lac were very close to 3:4 and 4:5 resonances. Rich modulation frequency patterns with multiplets of the same modulation frequency were revealed in extended, accurate observations of some Blazhko stars \citep[MW Lyr and V1127 Aql in][]{mwI,aql}. In these cases the dominant modulation corresponded to the shortest frequency component of the multiplets.
However, the examples of V2, V14 and CZ Lac show that in some Blazhko stars, from time to time, different components of a hidden modulation frequency multiplet may emerge. If these modulations correspond to elements of series of equidistant-spacing multiplets, their frequency ratios are close to fractions of small integer numbers. 
This indicates that the observed modulation-frequency values may not always correspond to the `base frequency', i.e. the shortest frequency component of a multiplet. One must be cautious in interpreting the data.

\subsection{Pulsation and modulation periods and period changes}

\begin{figure}
\includegraphics[angle=0,width=7.2cm]{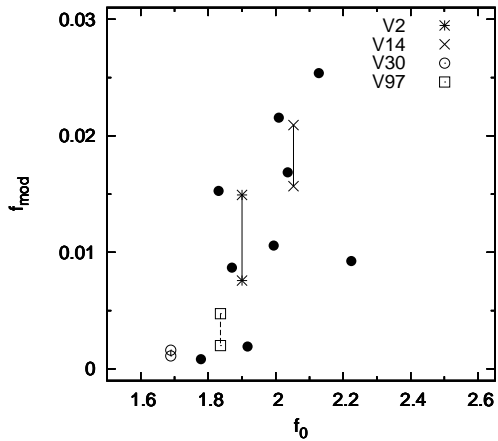}
\caption{The modulation versus pulsation frequencies of the Blazhko stars detected in M5 are plotted. The significantly different modulation periods observed at different epochs for V2 and V14 are also shown. The possible alias solutions for the modulation periods of V30 and V97 are plotted to indicate the uncertainties arising from incomplete data coverage. The formal errors of the frequencies do not exceed the symbols' size.
 }
\label{mod}
\end{figure}

Using a large sample of Blazhko variables, including stars from the galactic field, globular clusters, the galactic bulge, the LMC and the Sagittarius-dwarf galaxy, \cite{acta} showed that the larger the pulsation frequency of a Blazhko star is, the larger its modulation frequency can be. 
The modulation versus pulsation frequencies of the M5 Blazhko variables are plotted in Fig.~\ref{mod},
and it appears that a similar trend might occur in M5. However, the ranges of the observed pulsation and modulation frequencies are much smaller in M5 than in the sample investigated by \cite{acta}.
Most probably, if a relation between the pulsation and the modulation frequencies indeed exists, its actual form strongly depends on the global parameters of the stellar population studied (e.g, luminosity, chemical composition, age, etc.).

\begin{table*}
\caption{Pulsation- and modulation-period variations of Blazhko RRab stars}
 \label{bllong}
  \begin{tabular}{lccccrrr}
\hline
{star}
& $P_{0}$ [d]
& $\mid {\mathrm{d} P_{0}} \mid$ [$10^{-5}$d]
& $P_{\mathrm{Bl}}$ [d]
& $\mid {\mathrm{d} P_{\mathrm{Bl}}} \mid$ [d]
& ${{{\rm d}{P_{\mathrm{Bl}}}\over{P_{\mathrm{Bl}}}}} / {{{\rm d}{P_{0}}\over{P_{0}}}} $ 
&${{\rm d}{P_{\mathrm{Bl}}}\over{{\rm d} {P_{0}}}}$[$10^{4}$]
&${{\rm d}{f_{\mathrm{Bl}}}\over{\rm d}f_{0}}$ \\
\hline
RR Gem$^{a}$ &  0.3973 & 4.0&  7.2& 0.07&    96&  0.2  &  5.26\\
DM Cyg$^{b}$ &  0.4199 & 0.5& 10.6& 0.07&  -554& -1.4  &-21.94\\        
RW Dra$^{c}$ &  0.4429 &11.0& 41.8& 0.5&    -48& -0.5  & -0.51\\      
M5 V4$^{d}$  &  0.4496 & 5.3& 110 & 3.2&   -250& -6.0  & -1.04\\      
XZ Cyg$^{e}$ &  0.4665 &15.0&  58 & 1.0&    -54& -0.7  & -0.43\\      
RV UMa$^{f}$ &  0.4681 & 0.5&  90 & 0.5&   -519&-10.0  & -2.70\\       
XZ Dra$^{g}$ &  0.4765 & 5.0&  76 & 4.0&    505&  8.0  &  3.19\\       
M5 V58$^{d}$ &  0.4913 & 0.5&  59 & 0.4&   -666& -8.0  & -5.53\\        
M5 V63$^{d}$ &  0.4977 & 0.4&  46.5&0.2&   -535& -5.0  & -5.73\\       
M5 V2$^{d}$  &  0.5263 & 1.7& 133 & 6.8&  -1582&-40.0  & -6.26\\       
M5 V56$^{d}$ &  0.5347 & 0.5& 115 & 0.9&    840& 18.0  &  3.92\\        
M5 V8$^{d}$  &  0.5462 & 0.6& 65.5&2.0&  -2790& -33.0 & -23.3\\        
M5 V72$^{d}$ &  0.5623 & 0.6& 1200&31 &  -2414&-520.0 & -1.12\\      
\hline
\multicolumn{8}{l}{\footnotesize{References:$^{a}$ \citet{rrgII}; $^{b}$ \citet{dmcyg}; $^{c}$ Szeidl private comm.; }}\\
\multicolumn{8}{l}{\footnotesize{$^{d}$ this paper; $^{e}$ \cite{xzc}; $^{f}$ \cite{rvuma}; $^{g}$ \cite{xzdra}.}}\\
\end{tabular}
\end{table*}

The long time-base of the observations of the M5 variables also made it possible to determine the modulation properties at different epochs for some variables. Significant changes in the modulation properties were already detected in RR Lyr \citep{rrl}, RR Gem \citep{rrgII}, XZ Cyg \citep{xzc} and CZ Lac \citep{czl}. In these stars the strengths of the amplitude and the phase modulations  as well as the pulsation and the modulation periods vary. These changes are probably connected either to multi-periodicity of the modulation or to abrupt and continuous changes of the pulsation period.  In Table~\ref{bllong} we summarize the detected changes in the pulsation and the modulation periods for all those variables, which have data at least for two different epochs. We note, however, that in those cases when the observed period changes have been determined for several epochs (e.g. XZ Dra, XZ Cyg, RW Dra, RV UMa), no strict relation between the pulsation- and the modulation-period changes of the variables holds. The connection between the period changes reflects some tendency rather than an exact relation. Therefore, results obtained from two-epoch data have to be taken with caution. 

The columns of Table~\ref{bllong} give the pulsation and the modulation periods, the absolute values of their changes calculated from the periods given in Table~\ref{bltab} or taken from the literature, the normalized relative period variations, and the relative period and frequency variations for seven M5 and six field RRab stars, listed in order of increasing pulsation period.
Examining these data we failed to find any regular pattern, but some general trends are evident. First, it seems that in general, there is an anticorrelation between the directions of the pulsation and the modulation period changes, i.e. a negative sign of ${{\rm d}P_{\mathrm{Bl}}/{\rm d}P_0}$ is more common than a positive one. Secondly, it can be noticed that the relative period variations (${{\rm d}{P_{\mathrm{Bl}}}/{{\rm d} {P_{0}}}}$) and their normalized values tend to have larger absolute values for variables with longer pulsation periods than for shorter period ones. 

The pulsation-period variation of one of the Blazhko stars, V56, can be described with a long, periodic/cyclic variation. This gives a unique possibility to test the binary origin of the long-period $O-C$ variation. If this were caused by the light-time effect the pulsation and modulation periods should have to show similar variation, because $\Delta P/P$  is the function of the elements of the orbit for any periodic signal \citep{co71}. The observed normalised relative period variation of the pulsation and the modulation periods of V56 is, however, much larger than 1, it is 840 contradicting the orbital origin of the cyclic, long-period variation.  However, keeping in mind that random variations in the modulation periods may override the supposed synchronous variations of the pulsation and the modulation periods, this result is not in fact conclusive.

\section{Conclusions}

Although Blazhko stars are usually numerous in globular clusters, which are rich in RR Lyrae stars, we hardly know anything about their properties because of a lack of data suitable for studying them. Results have been restricted to deriving the Blazhko periods, but even they have been successfully obtained only in a few cases, e.g. in M3 \citep{m3}. In this paper the RRab population of M5 are reviewed using all the available photometric data of the variables, in order to identify the Blazhko stars and to study their behaviour. 
Modulation periods were determined for 13 out of the 20 Blazhko stars that were identified.
The long time base of the observations also made it possible to record changes in the observed properties of some of the Blazhko stars.

The most important results of this study can be summarized as follows:

\begin{itemize}
\item{Blazhko stars tend to be located in the short-period, low-luminosity region of the fundamental-mode instability strip. Their $B-V$s overlap with the domain occupied by the first-overtone pulsators.}
\item{The pulsation-period variation of most of the Blazhko stars is slightly or significantly irregular.}
\item{There is no clear connection between the pulsation and the modulation period changes of the Blazhko stars, both parallel and antiparallel changes are detected.}
\end{itemize}

Finally, we note that the possibilities of studying globular-cluster Blazhko stars are still unexploited. Although these studies would need relatively large observational efforts, the homogeneity of such samples would be helpful for determining what properties of the RR Lyrae stars predestine them to become Blazhko variables.

\section*{Acknowledgments}
The constructive, helpful comments of the referee, Katrien Kolenberg are much appreciated.
The financial support of OTKA grant K-068626 is acknowledged. 
C. Clement thanks the Natural Science and Engineering
Council of Canada for financial support.
Zs. H. thanks the `Lend\"ulet' program of the Hungarian Academy of Sciences for supporting his work.

\label{lastpage}
\end{document}